\documentclass[sn-mathphys-ay]{sn-jnl}

\usepackage{graphicx}%
\usepackage{multirow}%
\usepackage{amsmath,amssymb,amsfonts}%
\usepackage{amsthm}%
\usepackage{mathrsfs}%
\usepackage{subcaption}
\usepackage[title]{appendix}%
\usepackage{xcolor}%
\usepackage{textcomp}%
\usepackage{manyfoot}%
\usepackage{booktabs}%
\usepackage{algorithm}%
\usepackage{algorithmicx}%
\usepackage{algpseudocode}%
\usepackage{listings}%
\usepackage{standalone}%
\usepackage{bm}
\usepackage{multirow}   
\usepackage{array}      
\usepackage{booktabs}
\usepackage{tikz}
\usepackage{fix-cm}
\usepackage{lmodern}
\usepackage{rotating}
\usepackage{graphicx}
\usepackage{float}
\usepackage{setspace}
\usepackage{placeins}
\usepackage{colortbl}
\usepackage{xcolor}
\usepackage{hhline}
\usepackage{hyperref}
\usepackage{natbib}
\usepackage{changepage}
\usepackage[utf8]{inputenc}
\usepackage{booktabs}
\usepackage{multirow}
\usepackage{tabularx} 
\usepackage{needspace}  
\usepackage{makecell}

\theoremstyle{plain}

\theoremstyle{definition}

\theoremstyle{remark}

\begin{document}

\title[Scalable Variational Inference for Multinomial Probit Models under Large Choice Sets and Sample Sizes]{Scalable Variational Inference for Multinomial Probit Models under Large Choice Sets and Sample Sizes}


\author[1]{\fnm{Gyeongjun} \sur{Kim}}\email{kkjunn7033@cau.ac.kr}

\author[1]{\fnm{Yeseul} \sur{Kang}}\email{dptmf0307@cau.ac.kr}

\author[2]{\fnm{Lucas} \sur{Kock}}\email{lucas.kock@nus.edu.sg}

\author[3]{\fnm{Prateek} \sur{Bansal}}\email{prateekb@nus.edu.sg}

\author[1,4]{\fnm{Keemin} \sur{Sohn}}\email{kmsohn@cau.ac.kr}
\equalcont{Corresponding author.}

\affil[1]{\orgdiv{Department of Urban Engineering}, \orgname{Chung Ang University}, \country{Korea}}

\affil[2]{\orgdiv{Department of Statistics and Data Science}, \orgname{National University of Singapore}, \country{Singapore}}

\affil[3]{\orgdiv{Department of Civil and Environmental Engineering}, \orgname{National University of Singapore}, \country{Singapore}}

\affil[4]{\orgdiv{Department of Smart City}, \orgname{Chung Ang University}, \country{Korea}}

\abstract{The multinomial probit (MNP) model is widely used to analyze categorical outcomes due to its ability to capture flexible substitution patterns among alternatives. Conventional likelihood-based and Markov chain Monte Carlo (MCMC) estimators become computationally prohibitive in high-dimensional choice settings. This study introduces a fast and accurate conditional variational inference (CVI) approach to calibrate MNP model parameters, which is scalable to large samples and large choice sets. A flexible variational distribution on correlated latent utilities is defined using neural embeddings, and a reparameterization trick is used to ensure the positive definiteness of the resulting covariance matrix. The resulting CVI estimator is similar to a variational autoencoder, with the variational model being the encoder and the MNP's data generating process being the decoder. Straight-through-estimation and Gumbel-SoftMax approximation are adopted for the `argmax’ operation to select an alternative with the highest latent utility. This eliminates the need to sample from high-dimensional truncated Gaussian distributions, significantly reducing computational costs as the number of alternatives grows. The proposed method achieves parameter recovery comparable to MCMC.  It can calibrate MNP parameters with 20 alternatives and one million observations in approximately 28 minutes — roughly 36 times faster and more accurate than the existing benchmarks in recovering model parameters. }

\keywords{Multinomial Probit model, Variational inference, Neural embedding, Large choice set}



\maketitle

\section{Introduction}\label{sec1}

Discrete choice models are extensively employed across various subfields of applied economics, including agriculture, health, and transportation \citep{bib31}. Models from the generalized extreme value (GEV) family, particularly the multinomial and nested logit, are widely used due to their closed-form expressions for choice probabilities \citep{bib21, bib22}. However, this analytical convenience comes at the cost of imposing restrictive substitution patterns among alternatives. In contrast, the multinomial probit (MNP) model offers greater flexibility in capturing real-world decision-making by allowing for unrestricted correlation among alternative-specific latent utilities \citep{bib8, bib9}.   

Choice probabilities in MNP involve the computation of the cumulative density function of a multivariate Gaussian distribution. This makes likelihood-based estimation challenging because closed-form choice probability expressions are generally unavailable for typical choice dimensions. The likelihood-based estimators approximate the choice probability integrals via simulation \citep{bib31}, but estimation becomes computationally prohibitive as the size of the choice set increases. An alternative approach is Markov Chain Monte Carlo (MCMC), which avoids the direct computation of choice probabilities by treating latent utilities as model parameters \citep{bib1, bib19, bib20, bib35, bib38}. Since MCMC bypasses the need for solving the high-dimensional integral by relying on iterative sampling from truncated Gaussian distributions, it is often favored over the likelihood-based estimator. However, MCMC also struggles to scale for large choice sets because the dimensionality of the truncated Gaussian distribution increases with the number of alternatives, resulting in excessively high computational costs. 

Variational inference \citep[VI;~][]{BleKucMca2017,OrmWan2010} has emerged as a computationally efficient alternative to MCMC. The main idea of VI is to approximate the posterior distribution with a variational distribution by minimizing a measure of distance to the posterior distribution. 
Recently, there has been growing interest in applying VI to advanced choice models particularly for the mixed multinomial logit model \citep[MMNL;][]{McfTra2000}.
\citet{bib36} proposes the use of Laplace approximations and stochastic linear regression introducing mini-batches to scale approximations to large sample sizes.  \citet{BraMca2010} considers factorized Gaussian approximations. Building on this, \citet{KocTanBanNot2025} introduce skewness corrections to improve inference on individual-specific taste parameters. \citet{bib37} consider a fully factorized mean-field approximation and demonstrates significant improvements in predictive accuracy of VI methods under small and medium sized choice sets and sample sizes. A recent approach to scale VI for MMNL to large datasets, combining amortized VI with stochastic backpropagation, is given by \citet{bib27}. 

However, the application of VI to MNP remains challenging due to two reasons. First, the MNP model requires the covariance matrix of latent utilities to be positive definite, and the scale of the utilities must be normalized to ensure model identifiability. Common identifiability constraints include fixing a diagonal element of the variance-covariance matrix of utility differences or imposing restrictions on its trace \citep{bib20, bib38}. However, these constraints can introduce non-differentiability and non-convexity into the evidence lower bound (ELBO), leading to significant optimization challenges. Second, augmenting latent utilities makes it challenging to define a joint variational distribution over latent utilities, taste parameters, and variance-covariance components. Using a simplified variational distribution risks overlooking the complex correlations present in the true posterior distribution of these parameters.

For these reasons, research in developing VI for MNP is limited and often involves strong assumptions \citep[e.g.~][]{GirRog2006,FasDur2022}.
\cite{bib43}'s work on scalable VI for MNP remains state-of-the-art (LM-N approach, henceforth). They ensured parameter identifiability by transforming the covariance matrix of latent utilities into a spherical coordinate system, which implicitly enforces a trace restriction. To reduce the number of parameters and ensure positive definiteness, LM-N adopted a factor-structured covariance matrix, introducing the number of factors as an additional hyperparameter. To address the ELBO optimization issues, they use an exact conditional posterior distribution for the latent utilities as the conditional variational distribution while adopting Gaussian distributions for the remaining taste parameters and covariance parameters in the spherical system. As a result, the conditional variational distribution of latent utilities retain a truncated Gaussian distribution, limiting the computational efficiency of the approach. Significant computational gains in VI are mainly achieved through the subsampling of latent utilities in the stochastic gradient ascent algorithm. 

\begin{figure}[htbp]
\centering
\includegraphics[width=\columnwidth]{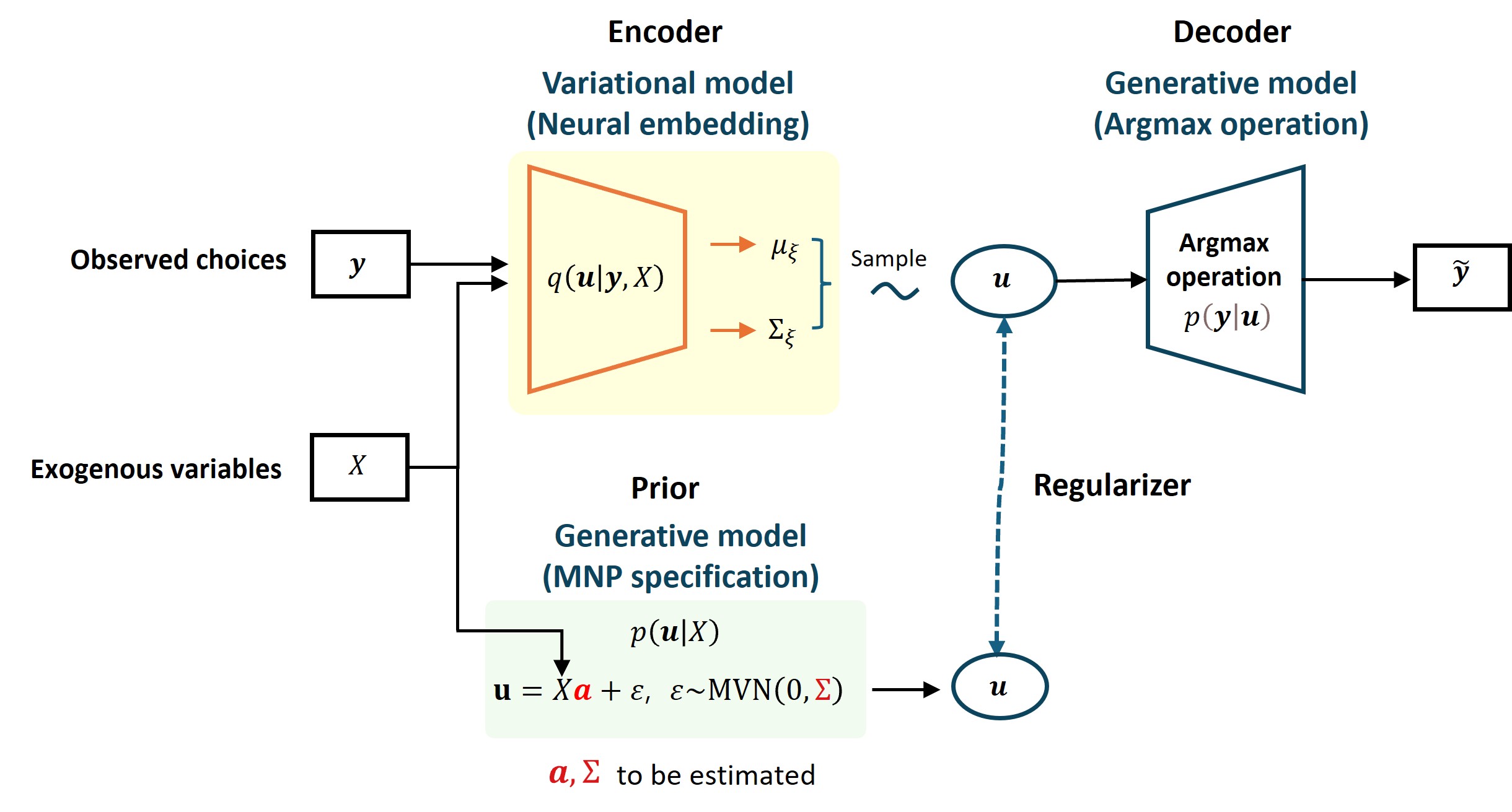}
\caption{An overview of the MNP parameter calibration framework.}
\label{Figure 1.}
\end{figure}

To address these challenges, this study presents a conditional VI (CVI) approach for calibrating MNP parameters. We briefly illustrate the core idea of CVI using abbreviated notation here. A comprehensive discussion is given later in Section~\ref{sec2}. In CVI, the variational distribution of correlated utilities $q(\boldsymbol{u}| \cdot)$  conditional on exogenous variables $(X)$ and observed choices $(\boldsymbol{y})$ is specified using flexible neural embeddings. Minimizing the Kullback-Leibler(KL) divergence between the conditional variational distribution of latent utilities $q(\boldsymbol{u}|\boldsymbol{y},X)$ and the true generative distribution $p(\boldsymbol{u}|\boldsymbol{y}, X)$  leads to a loss function with two terms. Minimizing the first term ensures an accurate reproduction of the observed choices, and the second loss term minimizes the deviation between the variational model $q(\boldsymbol{u}|\boldsymbol{y}, X)$ and the generative model $p(\boldsymbol{u}|X)$ for the latent utilities. The proposed calibration framework is illustrated in Fig. \ref{Figure 1.}. 
 
The key contributions of our novel calibration framework are as follows. First, the `argmax’ operation is used to select the alternative with the maximum latent utility $p(\boldsymbol{y}|\boldsymbol{U})$. Specifically, we use the straight-through estimation (STE) \citep{bib2} and the Gumbel-SoftMax trick \citep{bib15, bib41} to approximate the argmax function in a differentiable manner. This operation bypasses the need to sample from high-dimensional truncated Gaussian distributions, making the MNP parameter calibration scalable for large choice sets. Second,  we apply the reparameterization trick devised by \cite{bib18} on the covariance matrix in the variational model. Specifically, the covariance matrix is decomposed into a diagonal matrix with positive elements and a lower triangular matrix with diagonal elements normalized to one, ensuring its positive definiteness. The regularizer ensures that the positive definiteness also holds for prior covariance in the MNP model.  Third, the covariance matrix of the prior is normalized by imposing a trace restriction to circumvent identification issues. Fourth, the calibration framework takes a variational autoencoder structure, with the variational model as the encoder, the `argmax' operation as the decoder, and the prior as the regularizer.

We compare key aspects of our CVI approach with that of the LM-N approach. First, unlike LM-N, the proposed method bypasses truncated Gaussian distributions while satisfying necessary constraints on the covariance matrix. This strategy for omitting high-dimensional integrals offers substantial computational benefits and is transferable to other models involving latent variables and complex likelihoods. Second, similar to LM-N, our approach supports stochastic gradient descent, enabling scalability to large datasets through parallelized calibration. Third, unlike the LM-N approach, the proposed method is a calibration technique that provides point estimates rather than full posterior distributions. Since variational inference tends to underestimate posterior uncertainty, bootstrapping is commonly used to approximate uncertainty, which can also be easily applied within the proposed CVI framework. Fourth, in contrast to LM-N, our method leverages automatic differentiation, making it easier to implement using open-source machine learning frameworks and enhancing its broader applicability.  

The proposed CVI approach achieves parameter recovery comparable to MCMC as demonstrated in our simulation study. It can calibrate MNP parameters for datasets with 20 alternatives and one million observations in approximately 28 minutes, which is about 36 times faster and considerably more accurate in recovering model parameters than the LM-N approach. This efficiency largely stems from bypassing truncated Gaussian distributions. Combining this with stochastic gradient descent, the proposed method offers scalability in both sample size and choice set.

In the next section, we review MNP and the CVI framework. Section~\ref{sec3} introduces our novel calibration approach. The computational efficacy of our approach is illustrated on simulated data in Section~\ref{sec4} and a real data application to consumer choice data in Section~\ref{sec5}. Section~\ref{sec6} concludes.

\section{Conditional variational inference in the MNP model}\label{sec2}

\subsection{Generative Model for MNP} \label{2.1}

The MNP model assumes that the decision-maker chooses the alternative that maximizes the indirect utility. Let $u_{ij}$, $j=1,\dots,d$, denote the utility derived by decision-maker $i=1,\dots,n$ for the $j$-th choice out of a choice set with $d$ alternatives. The choice outcome vector $\boldsymbol{y_i}=(y_{i1},\dots,y_{id})$ is represented using a one-hot encoding scheme
\begin{align*}\
    y_{ij} = \begin{cases}
        1 \quad\text{if } j=\underset{j=1,\dots,d}{\arg\max} u_{ij}\\
        0 \quad\text{else},
    \end{cases}
\end{align*}
which we write compactly as $\boldsymbol{y_i}=\arg\max \boldsymbol{u_i}$. The unobserved latent utility vector $\boldsymbol{u_i}=\left(u_{i1},\dots,u_{id}\right)$ is assumed to be jointly Gaussian
\begin{equation*}
    \boldsymbol{u_i} = X_i \boldsymbol{a}  + \boldsymbol{\varepsilon_i} \quad \text{and,} \quad \boldsymbol{\varepsilon_i} \sim \mathrm{MVN}\left(\mathbf{0}, \Sigma\right),
\end{equation*}
where $X_i\in\mathbb{R}^{d\times p}$ is a known matrix of alternative- and individual-specific covariates, $\mathbf{a}\in\mathbb{R}^{p}$ is a vector of weights connecting the covariates with the utility, and $\boldsymbol{\varepsilon_i}\in\mathbb{R}^{d}$ is an idiosyncratic error vector including all unobserved factors influencing preferences with error or utility covariance matrix $\Sigma\in\mathbb{R}^{d\times d}$. The parameters to be learned are $\Sigma$ and $\boldsymbol{a}$, while pairs $(X_i,\boldsymbol{y}_i)$ for $i=1,\dots,n$ are observed.

The proposed calibration method is subject to the usual identification issue. Specifically, the utility covariance matrix $\Sigma$ is not identified. For example, multiplication with a scalar $\lambda>0$, $\Sigma \cong \lambda\Sigma$, does not change the marginal likelihood for $\boldsymbol{y_i}$ as absolute values of $\boldsymbol{u_i}$ remain unobserved. Various identification strategies have been introduced. For example, \cite{bib42} and \cite{bib20} fix a single element of the covariance matrix, \cite{bib38} imposes a trace restriction, and \cite{bib43} uses a spherical transformation combined with a trace restriction. 

Since only differences in utility matter, we redefine the MNP model in terms of the $(d-1)$-dimensional differenced utility vector and impose a trace restriction on the differenced covariance matrix $\Delta\Sigma$ of MNP \citep{bib38}. A differenced utility vector is obtained by subtracting the baseline alternative \(1\) from the utility of other alternatives
\begin{equation*}\
  {\Delta}\boldsymbol{u_i}
  =\bigl(u_{i2}-u_{i1},\,\dots,\,u_{id}-u_{i1}\bigr)^{\!\top}\!
  \in\mathbb{R}^{d-1}.
\end{equation*}
The observed choice is then
\begin{equation*}
  y_i({\Delta}\boldsymbol{u_i})=
  \begin{cases}
    j, & \displaystyle
         \max {\Delta}\boldsymbol{u_i}={\Delta}u_{ij}>0, \quad for \enspace i=1,\dots,n.\\[4pt]
    1, & \displaystyle
         \max {\Delta}\boldsymbol{u_i}<0,
  \end{cases}
\end{equation*}

The differenced utilities follow the linear-Gaussian model
\begin{equation*}
  \Delta\boldsymbol{u_i}
  ={\Delta}X_i\,\mathbf{a}+\boldsymbol{\varepsilon_i},
  \qquad
  \boldsymbol{\varepsilon_i}\sim\mathrm{MVN}\!\bigl(\mathbf{0},\,{\Delta}\Sigma\bigr),
\end{equation*}
where the differenced design matrix is

\begin{equation*}
  {\Delta}X_i=
  \begin{pmatrix}
    X_{i2}-X_{i1}\\
    \vdots\\
    X_{id}-X_{i1}
  \end{pmatrix}
  \in\mathbb{R}^{(d-1)\times p},
  \quad
  {\Delta}\Sigma\in\mathbb{R}^{(d-1)\times(d-1)}.
\end{equation*}

For clarity, let $\boldsymbol{\theta} = \{\mathbf{a},\,\Sigma\}$ denote the original parameters, and let $\boldsymbol{\theta}^\Delta = \{\mathbf{a},\,\Delta\Sigma\}$ denote the parameters in the differenced utility space.  

The joint distribution of the observed choice vector $(\boldsymbol{y}_i)$  and the latent utilities conditional on the model parameters can thus be expressed as 

\begin{align*}
    p(\boldsymbol{y}_i,\boldsymbol{u}_i \mid \boldsymbol{\theta}) 
    &= p(\boldsymbol{y}_i \mid \boldsymbol{u}_i)\, p(\boldsymbol{u}_i \mid \boldsymbol{\theta}, X_i), 
    \quad i=1,\dots,n, \\
    p\bigl(\boldsymbol{y}_i, \Delta \boldsymbol{u}_i \mid \boldsymbol{\theta}^{\Delta}\bigr) 
    &= p\bigl(\boldsymbol{y}_i \mid \Delta \boldsymbol{u}_i\bigr)\, p\bigl(\Delta \boldsymbol{u}_i \mid \boldsymbol{\theta}^{\Delta}, X_i\bigr),
    \quad i=1,\dots,n.
\end{align*}

However, marginalizing over the latent vector of unobserved utilities is challenging. This is in particular true when the number of choices $d$ is large.

\subsection{Conditional Variational Inference} \label{2.2}
We consider a CVI approach, where we consider a variational distribution $q_{\boldsymbol{\xi}}(\boldsymbol{u}_i\vert \boldsymbol{y}_i,X_i)$ to approximate the conditional distribution $p(\boldsymbol{u}_i\vert \boldsymbol{y}_i,X_i,\boldsymbol{\theta})$ 
. Here, $\boldsymbol{\xi}$ denotes a vector of learnable parameters independent of $i$ making the approximation amortized. 

We use neural embeddings to model the variational distribution, offering greater flexibility than the traditional mean-field approach \citep{bib32}. The effectiveness of neural embeddings in variational inference for discrete choice models has been previously demonstrated \citep{bib27,KimKanKee2025}.
Specifically, a neural network,  with neural parameters $\boldsymbol{\xi}$, takes observed data as input to generate the mean vector and variance-covariance matrix of a multivariate normal distribution over latent utilities. 

In the proposed CVI approach, the variational distribution $q_{\boldsymbol{\xi}}(\boldsymbol{u}_i\vert \boldsymbol{y}_i,X_i)$ is assumed to approximate the true posterior distribution $p(\boldsymbol{u}_i\vert \boldsymbol{y}_i,X_i,\boldsymbol{\theta})$. Thus, the loss function minimizes the Kullback-Leibler(KL) divergence between the conditional variational distribution of latent utilities $q_{\boldsymbol{\xi}}(\boldsymbol{u}_i\vert \boldsymbol{y}_i,X_i)$ and true posterior distribution $p(\boldsymbol{u}_i\vert \boldsymbol{y}_i,X_i,\boldsymbol{\theta})$, $
   \sum_{i=1}^n D_{KL}\left[{q_{\boldsymbol{\xi}}(\boldsymbol{u}_i\vert \boldsymbol{y}_i,X_i)}||{p(\boldsymbol{u}_i\vert \boldsymbol{y}_i,X_i,\boldsymbol{\theta})}\right]$.
Here, $ D_{KL}({q(x)}||{p(x)})=\mathbb{E}_{q(x)}[\log q(x) - \log p(x)]$, where $\mathbb{E}_{q(x)}[\cdot]$ denotes the expectation with respect to $q(x)$, is the Kullback-Leibler divergence. We can write 
\begin{equation*}
\begin{aligned}
  &D_{KL}\bigl(q_{\boldsymbol{\xi}}(\boldsymbol{u}_i\mid \boldsymbol{y}_i,X_i)
         \,\|\,p(\boldsymbol{u}_i\mid \boldsymbol{y}_i,X_i,\boldsymbol{\theta})\bigr]\\
 &= \mathbb{E}_{q_{\boldsymbol{\xi}}(\boldsymbol{u}_i\vert \boldsymbol{y}_i,X_i)}\left[\log q_{\boldsymbol{\xi}}(\boldsymbol{u}_i\vert \boldsymbol{y}_i,X_i)-\log \frac{p(\boldsymbol{u}_i,\boldsymbol{y}_i\vert X_i,\boldsymbol{\theta})}{p(\boldsymbol{y}_i\vert X_i,\boldsymbol{\theta})}\right]\\
&= \mathbb{E}_{q_{\boldsymbol{\xi}}(\boldsymbol{u}_i\mid \boldsymbol{y}_i,X_i)}
    \bigl[\log q_{\boldsymbol{\xi}}(\boldsymbol{u}_i\mid \boldsymbol{y}_i,X_i)
         -\log p(\boldsymbol{u}_i\mid \boldsymbol{y}_i,X_i,\boldsymbol{\theta})\bigr]\\
&= \log p(\boldsymbol{y}_i\mid X_i,\boldsymbol{\theta})
  - \mathbb{E}_{q_{\boldsymbol{\xi}}(\boldsymbol{u}_i\mid \boldsymbol{y}_i,X_i)}
    \bigl[\log q_{\boldsymbol{\xi}}(\boldsymbol{u}_i\mid \boldsymbol{y}_i,X_i)
         -\log p(\boldsymbol{u}_i,\boldsymbol{y}_i\mid X_i,\boldsymbol{\theta})\bigr)\\
&=\log p(\boldsymbol{y}_i\vert X_i,\boldsymbol{\theta}) - \mathbb{E}_{q_{\boldsymbol{\xi}}(\boldsymbol{u}_i\vert \boldsymbol{y}_i,X_i)}[\log p(\boldsymbol{y}_i\vert \boldsymbol{u}_i)]+  D_{KL}\left[q_{\boldsymbol{\xi}}(\boldsymbol{u}_i \vert \boldsymbol{y}_i,X_i)\, \|\, p(\boldsymbol{u}_i \vert \boldsymbol{\theta}, X_i)\right].
\end{aligned}
\end{equation*}
 The first term, the marginal log-likelihood $\log p(\boldsymbol{y}_i\vert X_i,\boldsymbol{\theta})$, is independent of $\boldsymbol{\xi}$. Therefore, the loss function $\mathcal{L}(\boldsymbol{\xi},\boldsymbol{\theta})$ in Eq. (\ref{eq:1}) minimizes the remaining two terms, which are the negative evidence lower bound (ELBO). The ELBO provides a tractable lower bound to the marginal log-likelihood

\begin{equation}
\begin{aligned}
\mathcal{L}(\boldsymbol{\xi},\boldsymbol{\theta}) 
&= - \mathcal{L}_{\text{ELBO}} \\
&= - \mathbb{E}_{q_{\boldsymbol{\xi}}(\boldsymbol{u}_i \vert \boldsymbol{y}_i,X_i)}\left[\log p(\boldsymbol{y}_i \vert \boldsymbol{u}_i)\right]
+  D_{KL}\left[q_{\boldsymbol{\xi}}(\boldsymbol{u}_i \vert \boldsymbol{y}_i,X_i)\, \|\, p(\boldsymbol{u}_i \vert \boldsymbol{\theta}, X_i)\right].
\end{aligned}
\label{eq:1}
\end{equation}

Eq.\eqref{eq:1} defines the target loss function to be minimized, where $\left\{\boldsymbol{\xi},\boldsymbol{\theta}\right\}$ denotes the set of parameters to be calibrated. Thus, our CVI approach results in a point estimator $\widehat{\theta}$ for $\theta$. However, it is straight forward to do uncertainty quantification with bootstrapping and we will illustrate that the low computational cost of our approach makes bootstrapping feasible even for large data sets in Section~\ref{sec4}. Minimizing the first term in Eq.\eqref{eq:1}, $-\mathbb{E}[\log p(\boldsymbol{y}_i\vert \boldsymbol{u}_i)]$, ensures that the utilities generated by the variational approximation $q_{\boldsymbol{\xi}}(\boldsymbol{u}_i\vert\boldsymbol{y}_i,X_i)$ accurately reproduce the observed choices. The second term $ D_{KL}({q_{\boldsymbol{\xi}}(\boldsymbol{u}_i\vert \boldsymbol{y}_i,X_i)}||{p(\boldsymbol{u_i}\vert\boldsymbol{\theta},X_i)})$ serves as a regularization to ensure that the variational model $q_{\boldsymbol{\xi}}(\boldsymbol{u}\vert\boldsymbol{y}_i,X_i)$ does not deviate significantly from the generative model $p(\boldsymbol{u_i}\vert\boldsymbol{\theta},X_i)$ for latent utilities. However, the first term involves the non-differentiable 'argmax' operator in $p(\boldsymbol{y}_i\vert \boldsymbol{u}_i)$, which prevents direct gradient-based optimization. To overcome this hurdle, we replace the 'argmax' operator with a smooth approximation, as detailed in Section~\ref{3.1}.

The loss function $\mathcal{L}(\boldsymbol{\xi},\boldsymbol{\theta})$ in Eq.(\ref{eq:1}) employs the original parameter vector \(\boldsymbol\theta = (\mathbf a,\;\Sigma)\) through \(p(\boldsymbol u_i\mid\boldsymbol\theta,X_i)\).   
To ensure parameter identifiability while minimizing the loss function, we impose a trace restriction on the prior covariance in the \((d-1)\)‐dimensional utility-difference space, as shown earlier in  Section \ref{2.1}.  When computing the KL regularizer, the second term in the loss function, we map the mean vector $\boldsymbol{\mu}_{\boldsymbol \xi}$ and covariance matrix $\Sigma_{\boldsymbol \xi}$ from the variational distribution $q_{\boldsymbol{\xi}}(\boldsymbol{u}_i \vert\boldsymbol{y}_i,X_i)$ into the differenced form $\{\Delta \boldsymbol{\mu}_{\boldsymbol \xi}, \Delta \Sigma_{ \boldsymbol \xi}\}$ to ensure alignment with the parameterization of the trace‐restricted prior. For this transformation, we use the difference operator \(C\) of dimension $(d-1) \times d$: 

\begin{equation*}
\begin{aligned}
  C = 
  \begin{pmatrix}
    -1 &  1 & 0 & \cdots & 0\\
    -1 &  0 & 1 & \cdots & 0\\
    \vdots &    &   & \ddots & \vdots\\
    -1 &  0 & 0 & \cdots & 1
  \end{pmatrix}_{(d-1) \times d}
\end{aligned}
\label{eq:11}
\end{equation*}

\begin{equation*}
\begin{aligned}
  q_{\boldsymbol \xi}^\Delta(\Delta \boldsymbol{u}_i)
  = \mathcal{N}\bigl(C\,\boldsymbol{\mu}_{\boldsymbol\xi},\,C\,\Sigma_{\boldsymbol\xi}\,C^{\top}\bigr)
\end{aligned}
\end{equation*}

The first term of the loss function, cross-entropy loss, does not require additional identification restriction as it only involves the parameters of the neural network $\boldsymbol{\xi}$.  The resulting loss function 
\begin{equation}\label{eq:2}
\begin{aligned}
\mathcal{L}^{\mathrm{id}}(\boldsymbol\xi,\boldsymbol\theta^{\Delta})
&=
-\,\mathbb{E}_{q_{\xi}(\boldsymbol u_i\mid\boldsymbol y_i,X_i)}
     \bigl[\log p(\boldsymbol y_i\mid\boldsymbol u_i)\bigr] \\[4pt]
&\quad
+\,D_{KL}\!\Bigl(
      q_{\boldsymbol\xi}^{\Delta}(\Delta\boldsymbol u_i\mid\boldsymbol y_i,X_i)
      \;\Big\|\;
      p(\Delta\boldsymbol u_i\mid\boldsymbol\theta^{\Delta},\Delta X_i)
    \Bigr).
\end{aligned}
\end{equation}
comprises two terms: the first retains the full \(d\)-dimensional utility space to compute the cross-entropy loss, while the second applies the KL divergence regularizer in the \((d-1)\)-dimensional differenced space under the trace-restricted prior. This adjustment in the loss function ensures that MNP parameters \(\boldsymbol\theta^{\Delta} = \{\boldsymbol a, \Delta\Sigma \}\)) are identified.


\subsection{Encoder–Decoder and Prior Specification}\label{2.3}

\begin{figure}[htbp]
\centering

\begin{subfigure}[t]{\columnwidth}
    \centering
    \includegraphics[width=\columnwidth]{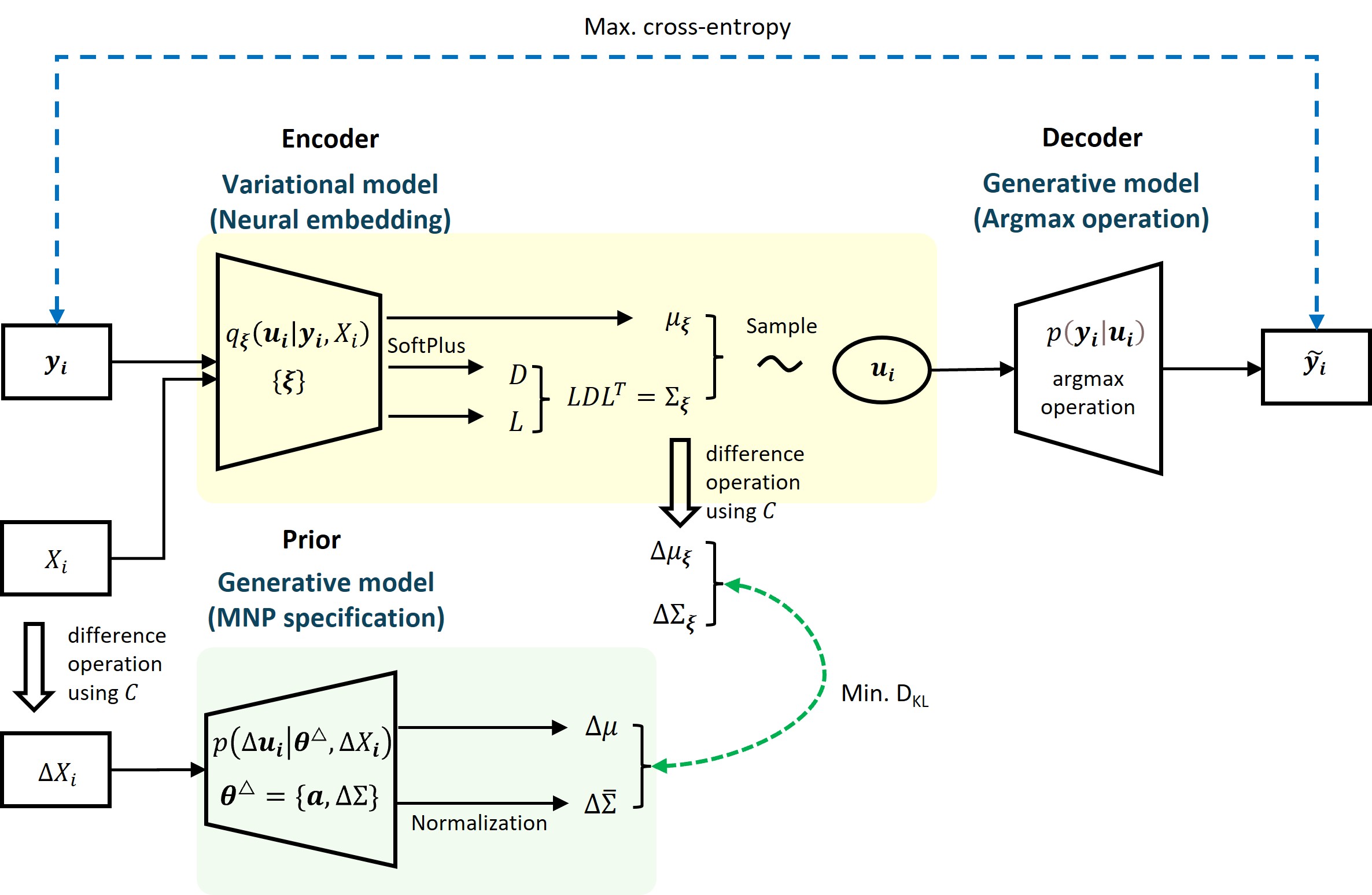}
    \caption{Calibration framework}
    \label{fig:2a}
\end{subfigure}

\vspace{1em}

\begin{subfigure}[t]{\columnwidth}
    \centering
    \includegraphics[width=10cm]{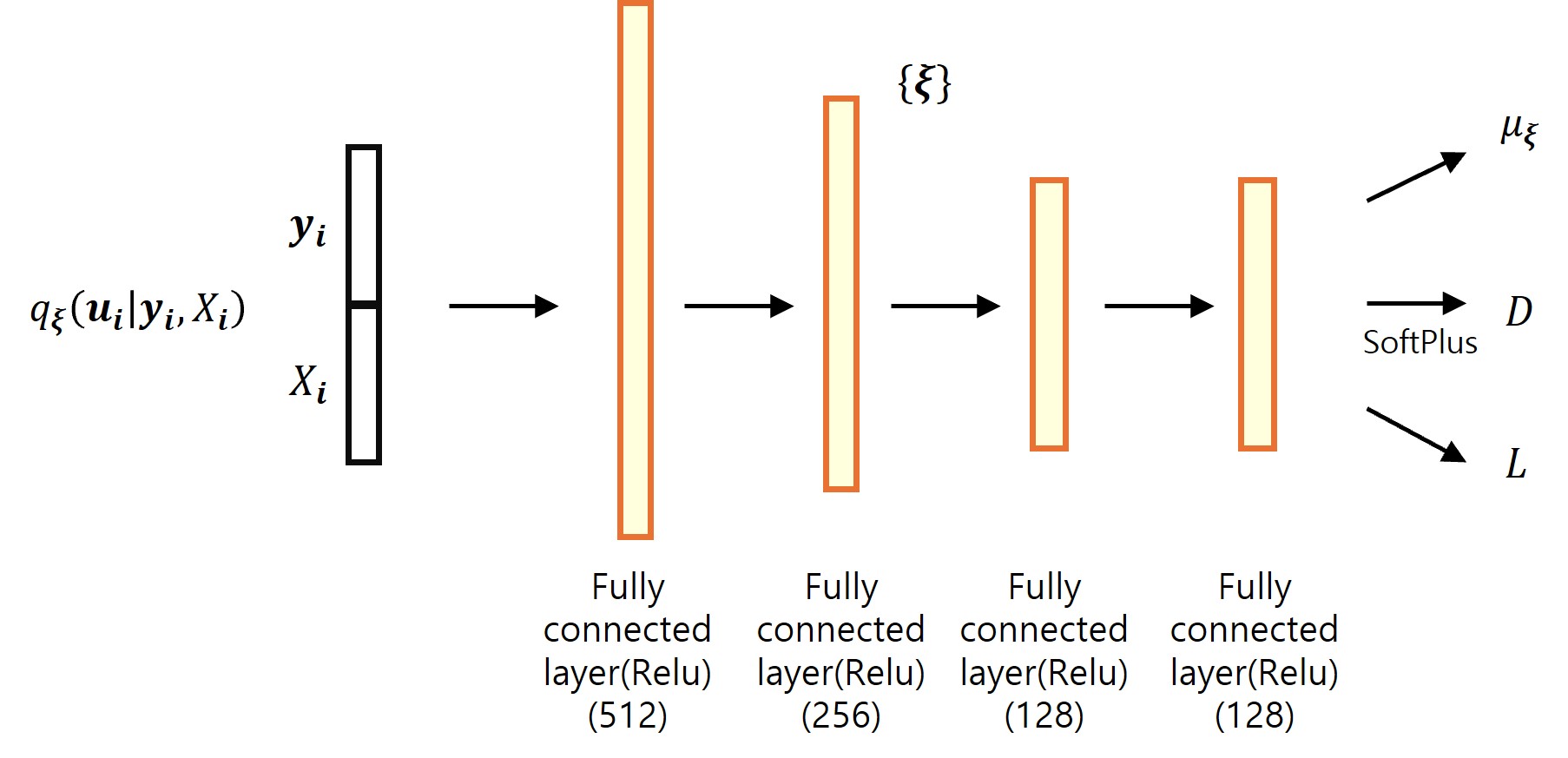}
    \caption{Network architecture for the variational model (encoder)}
    \label{fig:2b}
\end{subfigure}

\caption{Calibration framework for the proposed conditional variational inference approach}
\label{fig:2}

\end{figure}

Fig.~\ref{fig:2a} presents the calibration framework. The \textcolor{blue}{blue} dashed arrow corresponds to the cross–entropy term, while the \textcolor{green}{green} dashed arrow represents the KL-divergence-based regularizer in the loss function. Conceptually, the construction is consistent with a variational auto-encoder (VAE) architecture: the encoder $q_{\boldsymbol{\xi}}(\boldsymbol{u}_i\vert\boldsymbol{y}_i,X_i)$  amortises an intractable posterior, the decoder $p(\mathbf y_i\mid\mathbf u_i)$ is the deterministic 'argmax' operation from utilities to choices, and the MNP specification $p(\Delta\mathbf u_i\mid \boldsymbol{\theta}^{\Delta},\Delta X_i)$ serves as the prior in latent space. Each of these components is described below. 

The encoder $q_{\boldsymbol{\xi}}(\boldsymbol{u}_i\vert\boldsymbol{y}_i,X_i)$ approximates the  $p(\boldsymbol{u_i}| \boldsymbol{y}_i,X_i,\boldsymbol{\theta})$, $i=1,\dots,n$. We consider an amortized approach, where the vector of variational parameters $\boldsymbol{\xi}$ is shared across observations. To allow for the necessary flexibility, we use a deep neural network with a final distributional layer. In particular, $q_{\boldsymbol{\xi}}(\boldsymbol{u}_i\vert\boldsymbol{y}_i,X_i)$ is a Gaussian distribution with mean $\mu_{\boldsymbol{\xi}}(\boldsymbol{y}_i,X_i)$ and covariance matrix $\Sigma_{\boldsymbol{\xi}}(\boldsymbol{y}_i,X_i)$.  The distributional parameters $\mu_{\boldsymbol{\xi}}(\boldsymbol{y}_i,X_i)$ and $\Sigma_{\boldsymbol{\xi}}(\boldsymbol{y}_i,X_i)$ are the output of a fully connected deep neural network taking the observed choice outcome $\boldsymbol{y}_i$ and a vectorized covariate matrix $X_i$ as inputs, so that the sampling process of unobserved utilities within the variational model is given as
\begin{equation*}
    \boldsymbol{u}_i|\boldsymbol{y}_i, X_i \sim MVN \left( \boldsymbol{\mu}_{\boldsymbol \xi}(\boldsymbol{y}_i, X_i; \boldsymbol{\xi}), \Sigma_{\boldsymbol \xi}(\boldsymbol{y}_i, X_i; \boldsymbol{\xi}) \right).
\end{equation*}

 To ensure the positive-definiteness of the covariance matrix $\Sigma_{\boldsymbol{\xi}}(\boldsymbol{y}_i,X_i)$, the encoder outputs two components: a diagonal matrix $(D)$ with positive entries enforced via the SoftPlus activation and a lower triangular matrix $(L)$ with ones on the diagonal and unrestricted off-diagonal elements (see Fig.~\ref{fig:2a}). The covariance matrix is then computed as $\Sigma_{\boldsymbol \xi} = LDL^{T}$. For the KL-divergence term, the encoder covariance is converted into the $(d-1)$-dimensional trace-restricted space by applying the difference operator $C$, defined as $ \Delta{\Sigma_{\boldsymbol \xi}}=C\,{\Sigma_{\boldsymbol \xi}}\,C^{\top}$.

The decoder $p(\mathbf y_i\mid\mathbf u_i)$ deterministically selects the alternative with the highest utility, $y_i=\arg\max_{j\in\mathcal D_i}u_{ij}$. Since the `argmax' operator is non-differentiable, direct back-propagation through the cross-entropy term is not possible. Section~\ref{3.1} introduces a smooth surrogate function that combines the straight-through estimator with the Gumbel–Softmax trick, providing  gradients for stochastic-gradient optimization.

The reduced prior, ($p(\Delta \boldsymbol{u}_i\mid \boldsymbol{\theta}^\Delta,\Delta X_i)$) with \(\boldsymbol{\theta}^{\Delta} = (\boldsymbol{a},\,\Delta\Sigma)\), is the generative model for differenced utilities follows the MNP's formulation. To ensure parameter identifiability, we impose a trace restriction on $\Delta\Sigma$ such that \(\mathrm{trace}(\Delta\Sigma)=d-1\).  In practice, we scale the covariance matrix as follows:
\begin{equation}\label{eq:3}
\begin{aligned}
\Delta\bar{\Sigma} =\frac{d-1}{\mathrm{tr}(\Delta\Sigma)}\,\Delta\Sigma,
\end{aligned}
\end{equation}
which guarantees \(\mathrm{trace}(\Delta\bar{\Sigma})=d-1\). Although our MNP parameter are \(\boldsymbol\theta^{\Delta} = (\mathbf a,\,\Delta\Sigma)\), we always convert $\Delta\Sigma$ into $\Delta\bar{\Sigma}$ before computing the gradients of $\mathcal{L}^{\mathrm{id}}(\boldsymbol\xi,\boldsymbol\theta^{\Delta}$) during the model training.

\section{Scalable estimation and optimization}\label{sec3}

\subsection{Computation of the Loss Function} \label{3.1}

The loss function has two additive components 
\begin{equation*}
\begin{aligned}
  \mathcal{L}^{\mathrm{id}}(\boldsymbol{\xi},\boldsymbol\theta^{\Delta})
  &= \sum_{i=1}^n \Bigl[
       \mathcal{L}_1^{\mathrm{id}(i)}(\boldsymbol{\xi})
     + \mathcal{L}_2^{\mathrm{id}(i)}(\boldsymbol{\xi},\boldsymbol\theta^{\Delta})
     \Bigr], \\[4pt]
  \mathcal{L}_1^{\mathrm{id}(i)}(\boldsymbol{\xi})
  &= -\mathbb{E}_{q_{\boldsymbol{\xi}}(\boldsymbol{u}_i\mid \boldsymbol{y}_i,X_i)}
       \bigl[\log p(\boldsymbol{y}_i\mid \boldsymbol{u}_i)\bigr],
  \quad i = 1,\dots,n, \\[4pt]
  \mathcal{L}_2^{\mathrm{id}(i)}(\boldsymbol{\xi},\boldsymbol\theta^{\Delta})
  &= D_{KL}\!\bigl(
       q_{\boldsymbol{\xi}}^{\Delta}(\Delta\boldsymbol{u}_i\mid\boldsymbol{y}_i,X_i)
       \;\|\;
       p(\Delta\boldsymbol{u}_i\mid\boldsymbol\theta^{\Delta},\Delta X_i)
     \bigr),
  \quad i = 1,\dots,n.
\end{aligned}
\end{equation*}

\begin{figure}[!ht]
\centering
\includegraphics[width=0.35\columnwidth]{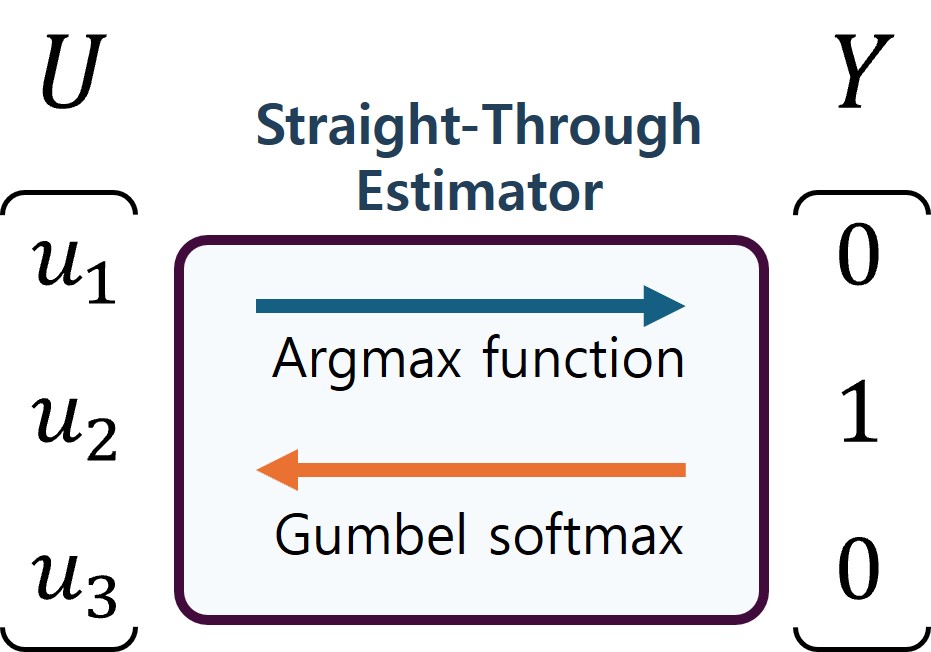}
\caption{STE incorporating the Gumbel softmax trick}
\label{fig:3}
\end{figure}

The first term $\mathcal{L}_1^{\mathrm{id}(i)}(\boldsymbol{\xi})$ is a cross-entropy expression for the reproduction of the choice data through the encoder model. However,  $p(\boldsymbol{y}_i\vert\boldsymbol{u}_i)$  corresponds to the `argmax’ operation that returns a one-hot vector indicating an alternative with maximum utility. `argmax' blocks gradient flow during backpropagation because it is non-differentiable, which is a common challenge in neural networks with discrete components.  To address this, \cite{bib33} proposed the straight-through estimator (STE), which applies `argmax' in the forward pass and `softmax' in the backward pass of the automatic differentiation. However, this mismatch can result in biases. Therefore, we replace the `softmax' with the Gumbel-Softmax \citep{bib15} in the backward pass to compute gradients, in which the $j$-th entry of $\boldsymbol{y}_i=\arg\max \boldsymbol{u}_i$ is replaced by
\begin{equation*}
  \widetilde{y}_{ij}
  = \frac{\exp\!\bigl((u_{ij} + g_{ij}) / \beta\bigr)}
         {\sum_{k=1}^d \exp\!\bigl((u_{ik} + g_{ik}) / \beta\bigr)},
  \quad j = 1, \ldots, d.
\end{equation*}
where $g_{ij} \sim \text{Gumbel}(0, 1)$ and $\beta$ is a hyperparameter. We write $\boldsymbol{\widetilde{y}}_i=\left(\widetilde{y}_{i1},\dots,\widetilde{y}_{id}\right)^\top$. 
Recently, \cite{bib41} illustrated the effectiveness of this approach for discrete latent variables. STE with the Gumbel-softmax trick is depicted in Fig. \ref{fig:3}.

The differentiable approximation $p(\boldsymbol{ \widetilde{{y}}_i}\vert \boldsymbol{u}_i)$ of the decoder $p(\boldsymbol{y}_i\vert \boldsymbol{u}_i)$ make $\mathcal{L}^{\text(i)}(\boldsymbol{\xi})$ differentiable. The reproduction loss term  can be evaluated approximately by sampling draws from the encoder model
\begin{equation*}
  \mathcal{L}_1^{\mathrm{id}(i)}(\boldsymbol{\xi})
  \approx -\frac{1}{L}\sum_{l=1}^L
    \boldsymbol{y}_i^\top \,\log \widetilde{\boldsymbol{y}}_i\bigl(\boldsymbol{u}_i^{(l)}\bigr),
\end{equation*}
where $\boldsymbol{u}_i^{(l)}\sim q_{\boldsymbol{\xi}}(\boldsymbol{u}_i\vert\boldsymbol{y}_i,X_i)$ are sample draws from the variational encoder model, $l=1,\dots,L$, $\boldsymbol{\widetilde{y}}_i$ is a continuous approximation of choice represented as a function of these sample draws, and $\boldsymbol{y}_i$ denotes the observed choice outcome. We make the dependence of $\boldsymbol{\widetilde{y}}_i$ on the samples explicit in the notation. 

The second term $\mathcal{L}^{\text{id}(i)}_2(\boldsymbol{\xi},\boldsymbol\theta^{\Delta})$ corresponds to the KL divergence between two multivariate Gaussian distributions and can thus be given in closed form
\begin{equation*}
\begin{aligned}
  \mathcal{L}^{\mathrm{id}(i)}_2(\boldsymbol{\xi},\boldsymbol\theta^{\Delta})
  &= \frac{1}{2}\Biggl[
       \log\frac{\det\bigl(\Delta\bar{\Sigma}\bigr)}
                {\det\bigl(\Delta\Sigma_{\boldsymbol{\xi}}(\boldsymbol{y}_i,X_i)\bigr)}
       \;-\;(d-1) \\[6pt]
  &\quad +\;\bigl(\Delta X_i\,\boldsymbol{a}
               - \Delta\mu_{\boldsymbol{\xi}}(\boldsymbol{y}_i,X_i)\bigr)^{\!\top}
            \Delta\bar{\Sigma}^{-1}
            \bigl(\Delta X_i\,\boldsymbol{a}
               - \Delta \boldsymbol{\mu}_{\boldsymbol{\xi}}(\boldsymbol{y}_i,X_i)\bigr) \\[6pt]
  &\quad +\;\mathrm{tr}\!\bigl(
       \Delta\bar{\Sigma}^{-1}
       \,\Delta\Sigma_{\boldsymbol{\xi}}(\boldsymbol{y}_i,X_i)
       \bigr)
    \Biggr].
\end{aligned}
\end{equation*}

Before evaluating the $\mathcal{L}^{\text{id}(i)}_2(\boldsymbol{\xi},\boldsymbol\theta^{\Delta})$, we normalize $\Delta{\Sigma}$  to $\Delta\bar{\Sigma}$, imposing the trace restriction shown in Eq.(\ref{eq:3}). This closed-form and differentiable expression simplifies the joint optimization of $\boldsymbol\theta^{\Delta}$ and $\boldsymbol{\xi}$. As a result, unlike MCMC, the proposed optimization framework does not require evaluating or sampling from $p(\Delta\boldsymbol{u}_i\vert\boldsymbol{\theta}^{\Delta},\Delta X_i, \boldsymbol{y}_i)$. 

\subsection{Stochastic optimization for large data sets} \label{3.2}
We will optimize the vector of model parameters $\boldsymbol{\nu}=(\boldsymbol{\xi},\boldsymbol\theta^{\Delta})^\top$ via stochastic gradient descent \citep{bib44}. If the number of observations $n$ is large, evaluation of $\mathcal{L}^{id}(\boldsymbol{\xi},\boldsymbol\theta^{\Delta})$ can be costly and a common technique to improve the computation time is to consider batch-wise optimization. Let $B$ be a subset of $\{1,\dots,n\}$ of cardinality $m$ chosen uniformly at random without replacement, inducing the set of indices of a minibatch of data. Then,
\begin{equation*}
  \widehat{\mathcal{L}}^{\mathrm{id}}(\boldsymbol{\xi},\boldsymbol\theta^{\Delta})
  = \frac{n}{m}
    \sum_{i\in B}\Bigl[
      \mathcal{L}_1^{\mathrm{id}(i)}(\boldsymbol{\xi})
    + \mathcal{L}_2^{\mathrm{id}(i)}(\boldsymbol{\xi},\boldsymbol\theta^{\Delta})
    \Bigr]
\end{equation*}
is an unbiased estimate of Eq.(\ref{eq:2}), so that $\boldsymbol{\nu}$ can be iteratively updated as $\boldsymbol{\nu}^{(t)}-\alpha^{(t)}\circ\nabla_{\boldsymbol{\nu}}\widehat{\mathcal{L}}^{\text{id}}(\boldsymbol{\xi}^{(t)},\boldsymbol\theta^{\Delta(t)})$, where $\alpha^{(t)}$ is an adaptive vector valued learning rate. We use the ADAM \citep{kingma2014adam} learning rate, and the gradients are obtained via automatic differentiation \citep{bib46}. Algorithm~\ref{algorithm} presents the full stochastic optimization procedure. This approach, especially when paired with GPU-accelerated parallel processing on batches, improved the computational efficiency of model training. This GPU-accelerated, batch-wise strategy makes our framework scalable to high-dimensional models in terms of the choice set size and the number of decision-makers. 

Appendix~\ref{appendix:A} presents the selected hyperparameter values. The hyperparameter $\beta$ in the Gumbel-Softmax distribution is systematically decreased across iterations to refine the discrete approximation \citep{bib15}.

\begin{algorithm}[H]
\caption{Scalable Variational Inference for MNP Model}
\label{algorithm}
\begin{algorithmic}
\setstretch{1.2}
\renewcommand{\algorithmicindent}{1em}
\State \textbf{Initialize} parameters $\boldsymbol{\nu}^{(0)}=(\boldsymbol{\xi}^{(0)},\boldsymbol{\theta}^{\Delta(0)})$
\State \textbf{Input} observed data $(\mathbf{y}, \mathbf{X})$
\State \textbf{Set} hyperparameters: learning rate $\alpha$, temperature $\beta$, mini-batch size $m$, number of utility samples ($L$) to approximate $\mathcal{L}_1^{\mathrm{id}(i)}(\boldsymbol{\xi})$ \textit{\small (see Appendix A)}
\State \textbf{Set} iteration counter $t \gets 0$

\While{not converged}
    \State \textbf{Sample} mini-batch $B$ of size $m$ from the observations
    \For{each $i \in B$}
        \State \textbf{Sample} latent utilities: $\boldsymbol{u}_{i}^{(l)}\sim q_{\boldsymbol{\xi}^{(t)}}(\boldsymbol{u}_i\vert\boldsymbol{y}_i,X_i)$ for $l = 1, \dots, L$
        \State \textbf{Compute:}  $\widehat{\mathcal{L}}^{\mathrm{id}(i)}_1(\boldsymbol{\xi}^{(t)})$ based on $\boldsymbol{u}_{i}^{(l)}, l=1,\dots,L$
        \State \textbf{Compute:}  $\widehat{\mathcal{L}}^{\text{id}(i)}_2(\boldsymbol{\xi}^{(t)},\boldsymbol{\theta}^{\Delta(t)})$
    \EndFor
    \State \textbf{Compute loss} $\widehat{\mathcal{L}}^{\text{id}}(\boldsymbol{\xi}^{(t)},\boldsymbol{\theta}^{\Delta(t)}) = \frac{n}{m} \sum_{i \in B} \left[\widehat{\mathcal{L}}_1^{\mathrm{id}(i)}\left(\boldsymbol{\xi}^{(t)}\right) + \widehat{\mathcal{L}}_2^{\mathrm{id}(i)}\left(\boldsymbol{\xi}^{(t)}, \boldsymbol{\theta}^{\Delta(t)}\right) \right]$
    \State \textbf{Compute gradient} $g^{(t)} = \nabla_{\boldsymbol{\nu}} \widehat{\mathcal{L}}^{\text{id}}(\boldsymbol{\xi}^{(t)},\boldsymbol{\theta}^{\Delta(t)})$
    \State \textbf{Update parameters} $\boldsymbol{\nu}^{(t+1)} = \boldsymbol{\nu}^{(t)} - \alpha^{(t)} \cdot g^{(t)}$
    \State $t \gets t + 1$
\EndWhile

\State \textbf{Return} optimized parameters $\boldsymbol{\nu}^{(t)}$
\end{algorithmic}
\end{algorithm}

\section{Simulation Study}\label{sec4}

\subsection{Set-up of the simulation study}

\paragraph{Reference Models}
To evaluate the performance of the proposed CVI approach, we benchmark it against two reference methods. First, we implement the MCMC approach \citep{bib47} to estimate the MNP model using Gibbs sampling. This involves drawing latent utilities from a truncated Gaussian distribution and iteratively updating taste parameters and covariances. Second, we adopt the VI based LM-N method of \cite{bib43}. This approach retains exact distributions for latent utilities (i.e., truncated Gaussian) and approximates other parameters with Gaussian distributions. Computational efficiency is further enhanced through stochastic gradient ascent with sub-sampling of latent utilities, allowing scalable inference for large samples and choice sets. We follow a consistent approach for a fair comparison and employ 10\% sub-sampling in our LM-N implementation for large-scale experiments with ten or more alternatives, while excluding sub-sampling in small-scale experiments with three alternatives.

\paragraph{Data Generating Process}
We evaluate the proposed approach against MCMC and LM-N on small-scale datasets with 3 alternatives and 5000 observations. To assess scalability, we also benchmark our CVI approach on larger datasets  -- 10 and 20 alternatives and sample sizes of 100,000, 500,000, and 1,000,000. MCMC is only estimated for the ten-alternative case, but is excluded for the twenty-alternative scenario due to its prohibitively high computational cost. True parameter values for the three-alternative case are: \\ 
\begin{align*}
X_i &= \begin{pmatrix} 
X_{i11} & 0 & 0 & 0 & X_{i12} \\
0 & X_{i21} & 0 & 0 & X_{i22} \\
0 & 0 & X_{i31} & X_{i32} & X_{i33}
\end{pmatrix}, \\
\Delta X_i &= \begin{pmatrix} 
-X_{i11} & X_{i21} & 0 & 0 & X_{i22} - X_{i12} \\
-X_{i11} & 0 & X_{i31} & X_{i32} & X_{i33} - X_{i12}
\end{pmatrix}, \\
\boldsymbol{a} &= \begin{pmatrix} 0.6 \\ 0.55 \\ 0.9 \\ -0.25 \\ 0.2 \end{pmatrix}, \\
\Delta \Sigma &= \begin{pmatrix} 0.89 & 0.31 \\ 0.31 & 1.11 \end{pmatrix}.
\end{align*}

The last entry of $\boldsymbol{a}$ is a generic parameter, i.e., all alternatives share this value. Each entry of the explanatory variable matrix $(X_i)$ was sampled from a standard uniform distribution. The choice outcome $(\mathbf{y})$ was determined by applying `argmax' operation on latent utilities, which were obtained by summing systematic utility and error sampled from a multivariate Gaussian distribution with mean 0 and covariance $\Delta\Sigma$. The same configuration applies to the large-scale dataset.

\paragraph{Comparison Metrics} 

To evaluate the performance of different models, we use four metrics: hit rate, log-score, Brier score, and root mean square error (RMSE). 

To evaluate the accuracy of choice prediction, we adopted the hit rate, which quantifies the proportion of cases in which the model’s most likely choice matches the observed choice. Given predicted choice vector \(\{\hat{\boldsymbol{y}}_i\}\), the hit rate is defined as
\begin{equation*}
    \text{hit-rate}
    = \frac{1}{N} \sum_{i=1}^{N} \mathbb{I}\bigl(\hat{\boldsymbol{y}}_i = \boldsymbol{y}_i\bigr),
\end{equation*}
A higher hit rate indicates better predictive accuracy.

To quantify probabilistic predictive accuracy, we adopt the log‐score \citep{GneRaf2007}, which quantifies the ability of the estimated model to assign high probabilities to observed choices. Given the estimated parameter set \(\hat{\bm{\theta}}^{\Delta}\), the log‐score is defined as
\begin{equation*}
    \text{log‐score}
    = \frac{1}{N} \sum_{i=1}^{N} \log p\bigl(\hat{y}_{ij}\mid \Delta X_i,\hat{\bm{\theta}}^{\Delta}\bigr),
\end{equation*}
where \(p(\hat{y}_{ij}\mid X_i,\hat{\bm{\theta}})\) represents the probability assigned to the observed choice \(y_{ij}\). A higher log‐score indicates better predictive accuracy.

To evaluate the distribution of the predicted choice probabilities, we used the Brier score \citep{GneRaf2007}: 
\begin{equation*}
    \text{Brier-score}
    = \frac{1}{N} \sum_{i=1}^{N}
       \left\| p\bigl(\hat{\boldsymbol{y}}_i\mid \Delta X_i,\hat{\boldsymbol{\theta}}^{\Delta}\bigr) - \boldsymbol{y}_i \right\|_2,
\end{equation*}
A lower Brier score indicates better calibration.

To evaluate parameter calibration accuracy, we adopt RMSE. It measures the average deviation between estimated and true parameter values. Given estimated parameter vector $\hat{\boldsymbol{\theta}}^{\Delta}$ and true values $\boldsymbol{\theta}^{\Delta}$, RMSE is defined as: 

\begin{equation*}
    \mathrm{RMSE}
    = \sqrt{\frac{1}{K} \sum_{k=1}^{K} (\hat\theta_k^{\Delta} - \theta_k^{\Delta})^2},
\end{equation*}

where $K$ is the number of parameters, $\hat{\theta}_k^{\Delta}$ is the posterior mean estimate of the $k^{th}$ parameter, and $\theta_k^{\Delta}$ is its true value. This metric summarizes the overall calibration accuracy by aggregating errors across all parameters.

\subsection{Results}

\begin{figure}[!ht]
\centering
\includegraphics[width=\columnwidth]{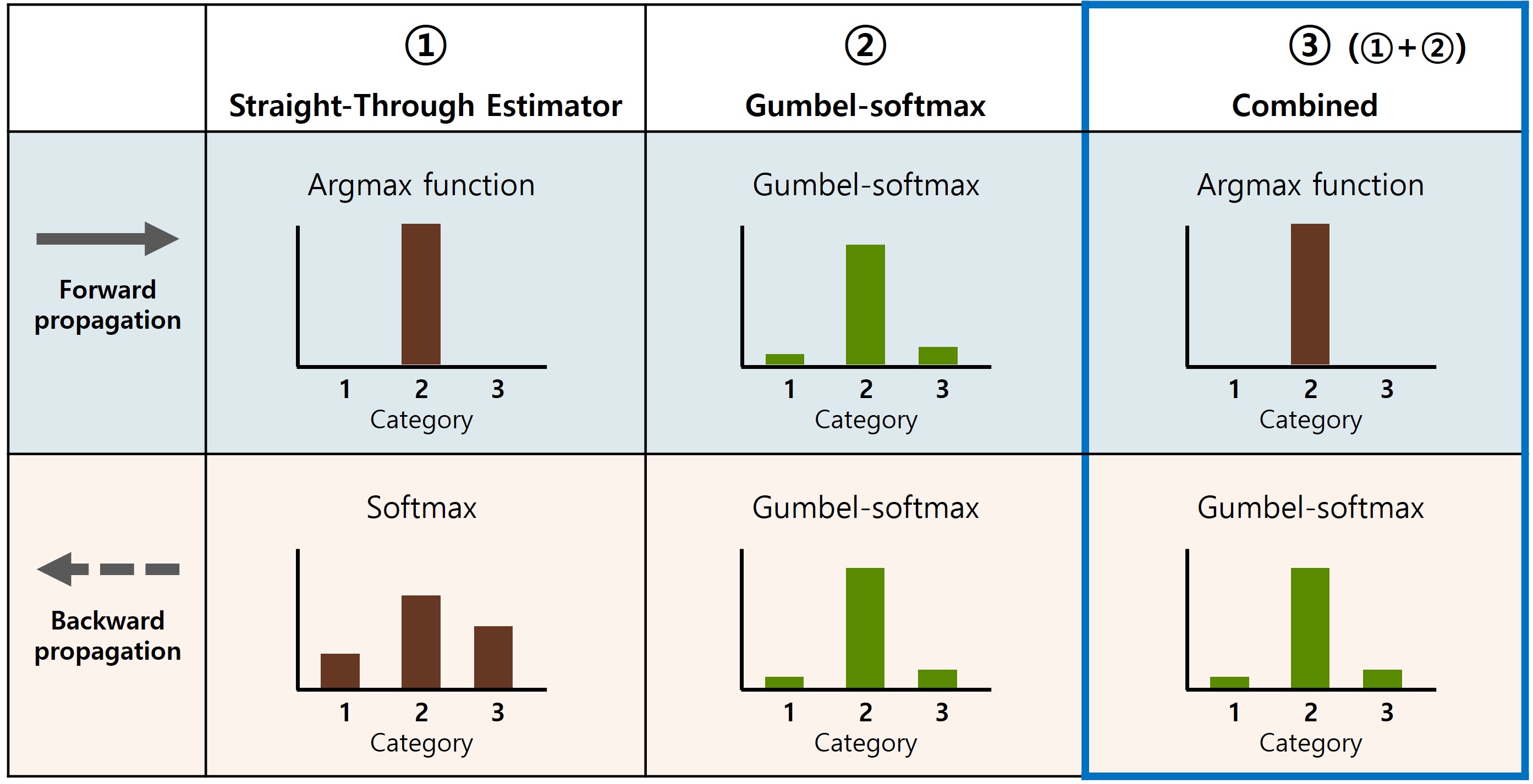}
\caption{Approximation methods for the ‘argmax’ operation}
\label{fig:4}
\end{figure}

\paragraph{Effect of different `argmax’ approximations} 
A key contribution of this study is to eliminate the need for costly evaluations of the cumulative distribution function of the multivariate Gaussian distribution through the `argmax' operation in the CVI framework. We examine the impact of three approximation strategies for deriving choice outcomes $\boldsymbol{y}_i$ from latent utilities $\boldsymbol{u_i}$, which are depicted in Fig. \ref{fig:4} . First, the STE \tikz[baseline=(char.base)]{\node[shape=circle,draw,fill=white,inner sep=1pt] (char) {1};} scheme directly applies the ‘argmax’ operation during forward propagation, generating a deterministic one-hot vector while using softmax in the backward pass. Second, the Gumbel-softmax technique \tikz[baseline=(char.base)]{\node[shape=circle,draw,fill=white,inner sep=1pt] (char) {2};} provides an approximation of the ‘argmax’ function during both forward and backward propagation, where random deviations and the temperature parameter are added to adjust the function’s kurtosis. The last method \tikz[baseline=(char.base)]{\node[shape=circle,draw,fill=white,inner sep=1pt] (char) {3};}  combines both techniques by employing ‘argmax’ in the forward pass and Gumbel-softmax in the backward pass, thereby maintaining differentiability while preserving choice consistency.

We compare these three approaches on a small-scale simulated dataset with three alternatives. 
Table \ref{tab1} shows that the combined approach \tikz[baseline=(char.base)]{\node[shape=circle,draw,fill=white,inner sep=1pt] (char) {3};} achieves the highest hit rate, indicating superior predictive accuracy. It also has better parameter recovery than other methods, as shown by the lowest RMSE value. Although the estimation time of the combined method is marginally higher than STE, its superior performance in parameter calibration and choice prediction makes it the most suitable approximation. We use the combined approach for the simulation study in subsequent subsections. 



\begin{table}[h]
\centering
\caption{Comparison of different ‘argmax’ approximations for small-scale data (three alternatives)}
\label{tab1}
\begin{tabular}{@{}cccccc cc@{}}
\toprule
\textbf{Model} & \textbf{Sample} & \textbf{Hit rate} & \textbf{Log score} & \textbf{Brier score} 
& \textbf{RMSE} 
& \textbf{\begin{tabular}[c]{@{}c@{}}Est. time\\ (min.)\end{tabular}} \\
\midrule

\multirow{2}{*}{\tikz[baseline=(char.base)]{\node[shape=circle,draw,inner sep=1pt] (char) {1};}~STE(softmax)} 
& In-sample   & 0.447 & -1.059 & 0.642 & \multirow{2}{*}{0.062} & \multirow{2}{*}{4.2} \\
& Out-of-sample  & 0.443 & -1.063 & 0.644 &                         &                      \\

\midrule

\multirow{2}{*}{\tikz[baseline=(char.base)]{\node[shape=circle,draw,inner sep=1pt] (char) {2};}~Gumbel-softmax} 
& In-sample   & 0.442 & -1.067 & 0.645 & \multirow{2}{*}{0.026} & \multirow{2}{*}{4.3} \\
& Out-of-sample  & 0.441 & -1.071 & 0.645 &                         &                      \\

\midrule

\multirow{2}{*}{\tikz[baseline=(char.base)]{\node[shape=circle,draw,inner sep=1pt] (char) {3};}~\textbf{Combined}} 
& In-sample   & 0.461 & -1.059 & 0.642 & \multirow{2}{*}{0.009} & \multirow{2}{*}{4.4} \\
& Out-of-sample  & 0.457 & -1.060 & 0.643 &                         &                      \\

\bottomrule
\end{tabular}
\end{table}

\paragraph{Convergence diagnostics}

\begin{figure}[H]
\centering
\includegraphics[width=\columnwidth]{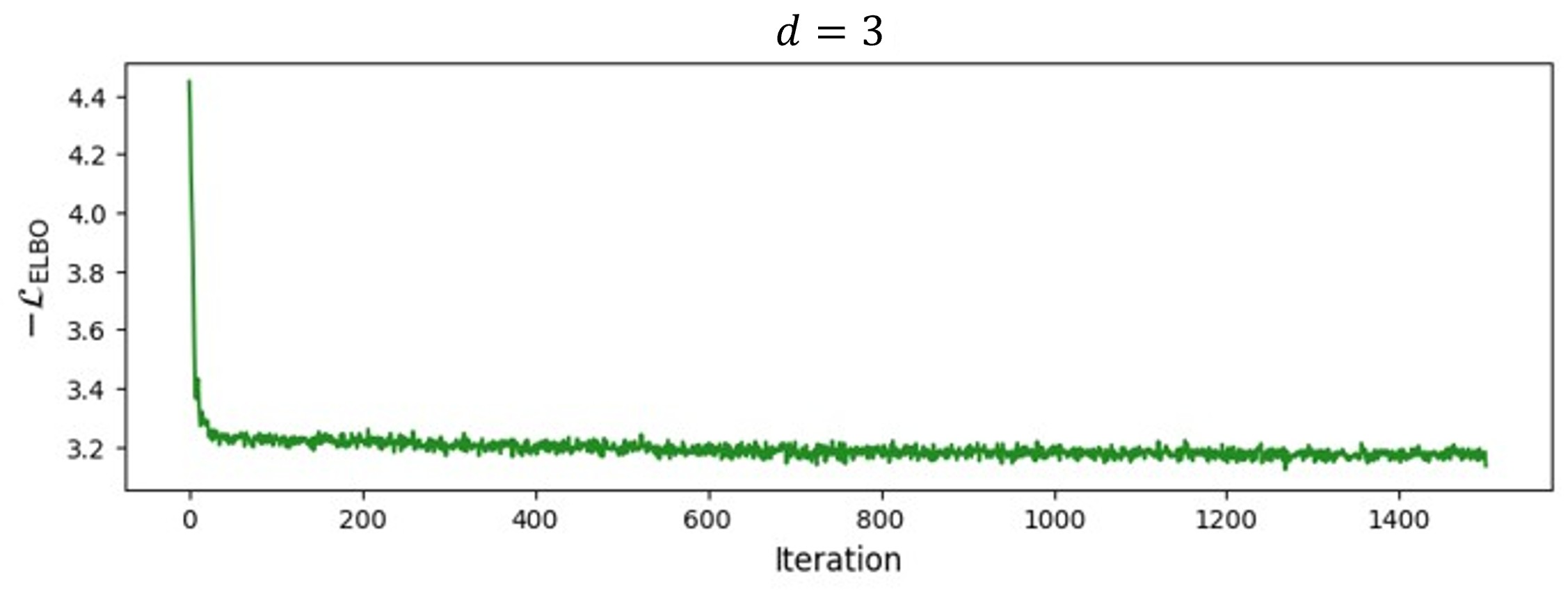}
\caption{Convergence of the proposed model for a small-scale dataset with three alternatives $(d=3)$}
\label{fig:5}
\end{figure}

\begin{figure}[H]
\centering
\includegraphics[width=\columnwidth]{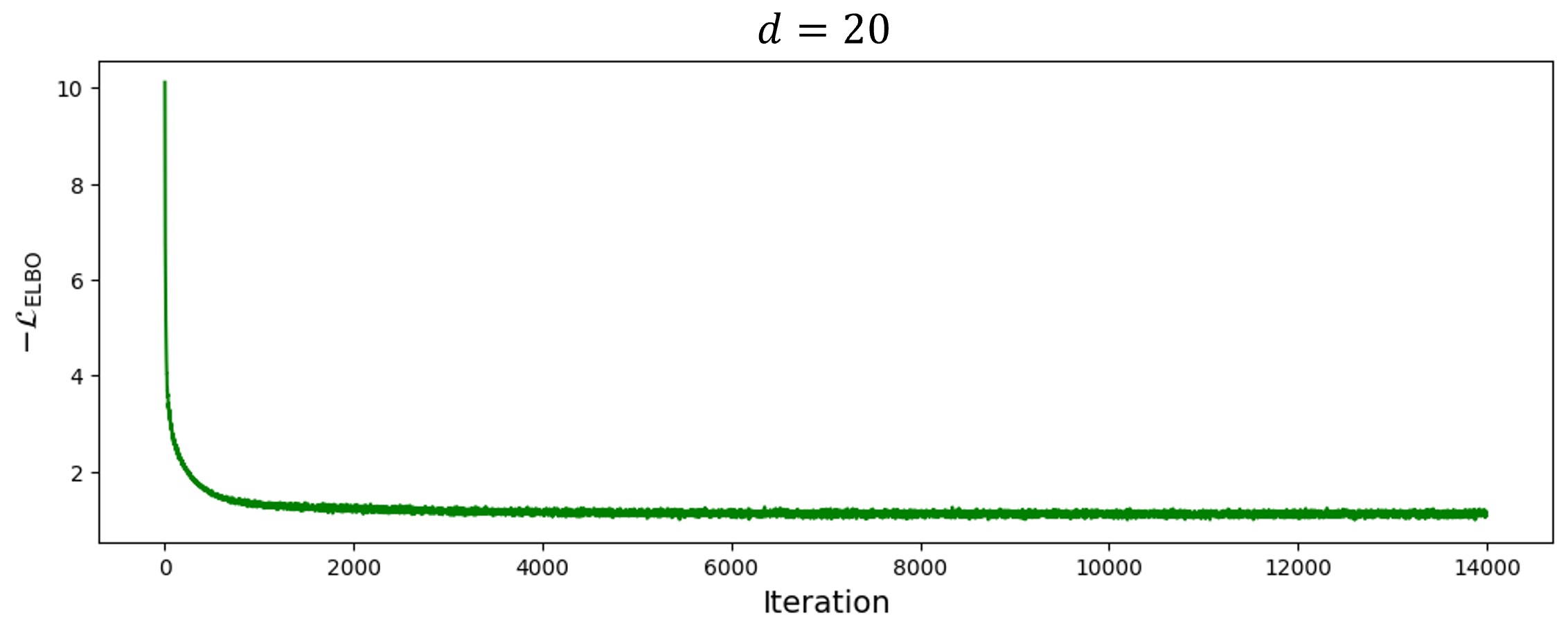}
\caption{Convergence of the proposed estimator for large-scale data (20 alternatives and 1 million observations)}
\label{fig:6}
\end{figure}

Fig. \ref{fig:5} presents the loss function over iterations of the proposed model for the small-scale simulated data with three alternatives. A rapid initial drop followed by stabilization indicates stable convergence behavior. A similar behaviour can be seen on the higher-dimensional data sets (Fig.~\ref{fig:6}).

\paragraph{Benchmarking against Reference Models} 

Table~\ref{tab2} compares the proposed CVI method with MCMC and LM–N on the small‐scale dataset. Predictive performance of the proposed method is nearly identical to the benchmarked methods in terms of hit rate, log score, and Brier score. However, there are differences in RMSE and estimation time. Specifically, the proposed method achieves the lowest RMSE (0.009), indicating the best parameter recovery. Although LM–N is marginally faster in estimation (3.8 min) than the proposed approach (4.4 min), it has the highest RMSE (0.066). MCMC attains a moderate RMSE (0.034) but incurs the longest estimation time (9.1 min). Thus, the proposed method offers the best parameter recovery without loss in predictive accuracy while maintaining computational efficiency closer to LM-N.  

\begin{table}[h]
\centering
\caption{Comparison of the proposed model against reference model on a small sample}
\label{tab2}
\begin{tabular}{@{}cccccc cc@{}}
\toprule
\textbf{Model} & \textbf{Sample} & \textbf{Hit rate} & \textbf{Log score} & \textbf{Brier score} 
& \textbf{RMSE} 
& \textbf{\begin{tabular}[c]{@{}c@{}}Est. time\\ (min.)\end{tabular}} \\
\midrule

\multirow{2}{*}{Proposed method}
& In-sample   & 0.432 & -1.064 & 0.644 & \multirow{2}{*}{0.009} & \multirow{2}{*}{4.4} \\
& Out-of-sample  & 0.423 & -1.074 & 0.652 &                         &                      \\

\midrule

\multirow{2}{*}{MCMC}
& In-sample   & 0.432 & -1.066 & 0.645 & \multirow{2}{*}{0.034} & \multirow{2}{*}{9.1} \\
& Out-of-sample  & 0.421 & -1.074 & 0.652 &                         &                      \\

\midrule

\multirow{2}{*}{LM-N}
& In-sample   & 0.432 & -1.064 & 0.644 & \multirow{2}{*}{0.066} & \multirow{2}{*}{3.8} \\
& Out-of-sample  & 0.427 & -1.072 & 0.649 &                         &                      \\

\bottomrule
\end{tabular}
\end{table}   

Table \ref{tab3} present posterior distributions of parameters for the three methods. The posterior distribution of parameters is obtained through sampling and optimization in MCMC and LM-N, respectively. In contrast, the proposed CVI approach only calibrates model parameters, resulting in point estimates. The sampling distribution of parameters is obtained using the bootstrap method. Table \ref{tab3} demonstrate that the point estimates of each parameter in the proposed approach are closer to the true values compared to other approaches, which aligns with the RMSE result. The proposed approach exhibits a marginally lower sampling variance for taste parameters $\boldsymbol{a}$ and covariance parameters ${\Delta \Sigma}$ compared to the posterior variance of MCMC and LM-N.

\clearpage

\clearpage
\begin{sidewaystable}
\centering
\FloatBarrier
\renewcommand{\arraystretch}{1.5}
\resizebox{1.05\textwidth}{!}{
\begin{tabular}{c c c c c c c c c c c c c c}
\toprule
\multirow{2}{*}{\textbf{Parameter}} & \multirow{2}{*}{\textbf{True}}
& \multicolumn{4}{c}{Proposed}
& \multicolumn{4}{c}{MCMC}
& \multicolumn{4}{c}{{LM-N}} \\ 
\cmidrule(lr){3-6} \cmidrule(lr){7-10} \cmidrule(lr){11-14}
& & \textbf{Mean} & \textbf{Std. dev.} & \multicolumn{2}{c}{\textbf{Credible interval}}
& \textbf{Mean} & \textbf{Std. dev.} & \multicolumn{2}{c}{\textbf{Credible interval}}
& \textbf{Mean} & \textbf{Std. dev.} & \multicolumn{2}{c}{\textbf{Credible interval}} \\  
\cmidrule(lr){5-6} \cmidrule(lr){9-10} \cmidrule(lr){13-14}
& & & & \textbf{Lower (2.5\%)} & \textbf{Upper (97.5\%)}
& & & \textbf{Lower (2.5\%)} & \textbf{Upper (97.5\%)}
& & & \textbf{Lower (2.5\%)} & \textbf{Upper (97.5\%)} \\  
\midrule
$\mathbf{a}_{1}$             & 0.6   & 0.594 & 0.046 & 0.502  & 0.687  & 0.576 & 0.053 & 0.469  & 0.682  & 0.535 & 0.033 & 0.468  & 0.603 \\
$\mathbf{a}_{2}$             & 0.55  & 0.554 & 0.045 & 0.463  & 0.646  & 0.516 & 0.047 & 0.422  & 0.611  & 0.504 & 0.037 & 0.429  & 0.578 \\
$\mathbf{a}_{3}$             & 0.9   & 0.901 & 0.049 & 0.802  & 1.001  & 0.858 & 0.054 & 0.749  & 0.967  & 0.809 & 0.058 & 0.693  & 0.925 \\
$\mathbf{a}_{4}$             & -0.25 & -0.233& 0.061 & -0.357 & -0.110 & -0.289& 0.067 & -0.424 & -0.155 & -0.285& 0.042 & -0.369 & -0.201 \\
$\mathbf{a}_{generic}$        & 0.2   & 0.199 & 0.036& 0.126  & 0.272  & 0.203 & 0.055 & 0.093  & 0.313  & 0.216 & 0.042 & 0.130  & 0.301 \\
$\bm{\Delta}\,\bm{\Sigma_{11}}$ & 0.89  & 0.909 & 0.062 & 0.785  & 1.033  & 0.845 & 0.086 & 0.674  & 1.022  & 0.799 & 0.112 & 0.575  & 1.024 \\
$\bm{\Delta}\,\bm{\Sigma_{22}}$ & 1.11  & 1.109& 0.062 & 0.985& 1.233& 1.150 & 0.086 & 0.977  & 1.325  & 1.20  & 0.112 & 0.975  & 1.424 \\
$\bm{\Delta}\,\bm{\Sigma_{12}}$ & 0.31  & 0.323 & 0.051 & 0.219  & 0.425& 0.321 & 0.076 & 0.167  & 0.474  & 0.351 & 0.093 & 0.164  & 0.538 \\

\bottomrule
\end{tabular}}
\begin{flushleft}
\footnotetext{* 95\% credible interval}
\end{flushleft}
\caption{Parameter estimates and credible intervals for the proposed and reference methods.}
\label{tab3}
\end{sidewaystable}
\clearpage

\clearpage

\paragraph{Scalability analysis}
To assess the scalability of the proposed CVI approach, we evaluated its computational efficiency and estimation accuracy against the MCMC and LM-N approach for choice sets with up to 20 alternatives and observation counts up to 1 million. MCMC was excluded for 20 alternatives due to its prohibitively high computational cost. Before discussing these results, the loss function convergence of the proposed approach is illustrated for the dataset with 20 alternatives and 1 million observations in Fig. \ref{fig:6}. 

In the LM-N approach, factorizing the covariance matrix is crucial to ensure estimation tractability by reducing the number of parameters. As detailed in \cite{bib43}, the covariance matrix is factorized as $BB^T + D^2$ , where $B$ is a $d \times p$ matrix and $D$ is a $d \times d$ diagonal matrix. Here, $p$ is a hyperparameter that controls the complexity of the factor structure. We evaluated the impact of hyperparameter $p$ on computational efficiency and estimation accuracy by testing two different settings: (i)  equal to the number of alternatives $(i.e., p=d)$, and (ii) half of the alternatives $(p=d/2)$. 

Table \ref{tab4} presents the comparative scalability results for $d=10$ and $d=20$ choice alternatives with 1 million observations.   For $d=10$, while all approaches yield nearly identical predictive performance in terms of hit rate, log-score, and Brier score, the differences are evident in parameter recovery (RMSE) and estimation time. While LM-N $(p=d/2)$ estimation takes 246 minutes, its estimation time is 18 times longer than the proposed method (13.5 minutes). Notably, the proposed method also marginally outperforms LM-N in terms of RMSE. MCMC yields the best RMSE score; however, its estimation time takes 4,504 minutes, which is prohibitively long. 

The differences in RMSE and estimation time between LM-N and the proposed method become more pronounced as the number of alternatives increases from 10 to 20. For 20 alternatives, LM-N ($p = d/2$) requires approximately 35 times more computation time than the proposed method (982 minutes vs.\ 28 minutes). In addition, the RMSE for LM-N ($p = d/2$) increases from 0.136 to 0.344 as the number of alternatives grows from 10 to 20, while the proposed method sees only a modest increase from 0.129 to 0.165. By avoiding the need to evaluate or sample from a truncated Gaussian distribution, the proposed method achieves substantial gains in computational efficiency while preserving strong parameter recovery performance for large choice sets.

Reducing the number of factors (\(p = d/2\)) helps alleviate some of the computational burden for LM-N, making it a more viable option for large-scale inference. However, even with this adjustment, LM-N (\(p = d/2\)) remains computationally inefficient, with a much higher RMSE than the proposed model. Results for 100,000 and 500,000 observations are reported in Appendix~\ref{appendix:B} and Appendix~\ref{appendix:C}. They exhibit similar relative performance patterns.

\begin{table}[h]
\centering
\caption{Comparison of models for $d=10$ and $d=20$ with 1,000,000 observations}
\label{tab4}
\renewcommand{\arraystretch}{1.3}
\begin{tabular}{@{}c c c c c c c c@{}}
\toprule
\textbf{$d$ / Obs.} & \textbf{Model} & \textbf{Sample} & \textbf{Hit rate} & \textbf{Log score} & \textbf{Brier score} & \textbf{RMSE} & \textbf{\begin{tabular}[c]{@{}c@{}}Est. time\\ (min.)\end{tabular}} \\
\midrule
\multirow{8}{*}{\shortstack{$d=10$\\1{,}000{,}000}} 
  & \multirow{2}{*}{\shortstack{Proposed\\method}} & In-sample   & 0.255 & -1.971 & 0.834 & \multirow{2}{*}{0.129} & \multirow{2}{*}{13.5}  \\
  &                                 & Out-of-sample  & 0.251 & -1.986 & 0.836 &                        &                       \\
  \cmidrule(lr){2-8}
  & \multirow{2}{*}{
  \shortstack{LM-N\\($p=d/2$)}}& In-sample   & 0.254 & -1.969 & 0.834 & \multirow{2}{*}{0.136} & \multirow{2}{*}{246}   \\
  &                                 & Out-of-sample  & 0.251 & -1.971 & 0.834 &                        &                       \\
  \cmidrule(lr){2-8}
  & \multirow{2}{*}{\shortstack{LM-N\\($p=d$)}}& In-sample   & 0.254 & -1.972 & 0.834 & \multirow{2}{*}{0.131} & \multirow{2}{*}{254}   \\
  &                                 & Out-of-sample  & 0.250 & -1.974 & 0.834 &                        &                       \\
  \cmidrule(lr){2-8}
  & \multirow{2}{*}{MCMC}            & In-sample   & 0.257 & -1.954 & 0.834 & \multirow{2}{*}{0.028} & \multirow{2}{*}{4504}  \\
  &                                 & Out-of-sample  & 0.255 & -1.958 & 0.836 &                        &                       \\
\midrule
\multirow{6}{*}{\shortstack{$d=20$\\1{,}000{,}000}} 
  & \multirow{2}{*}{\shortstack{Proposed\\method}} & In-sample   & 0.461 & -1.897 & 0.701 & \multirow{2}{*}{0.165} & \multirow{2}{*}{28}    \\
  &                                 & Out-of-sample  & 0.460 & -1.889 & 0.702 &                        &                       \\
  \cmidrule(lr){2-8}
  & \multirow{2}{*}{\shortstack{LM-N\\($p=d/2$)}}& In-sample   & 0.454 & -1.874 & 0.701 & \multirow{2}{*}{0.344} & \multirow{2}{*}{982}   \\
  &                                 & Out-of-sample  & 0.452 & -1.891 & 0.704 &                        &                       \\
  \cmidrule(lr){2-8}
  & \multirow{2}{*}{\shortstack{LM-N\\($p=d$)}}& In-sample   & 0.452 & -1.896 & 0.701 & \multirow{2}{*}{0.447} & \multirow{2}{*}{1016}  \\
  &                                 & Out-of-sample  & 0.451 & -1.898 & 0.704 &                        &                       \\
\bottomrule
\end{tabular}
\end{table}

\begin{figure}[H]
\centering
\begin{subfigure}[t]{\columnwidth}
    \centering
    \includegraphics[width=\linewidth]{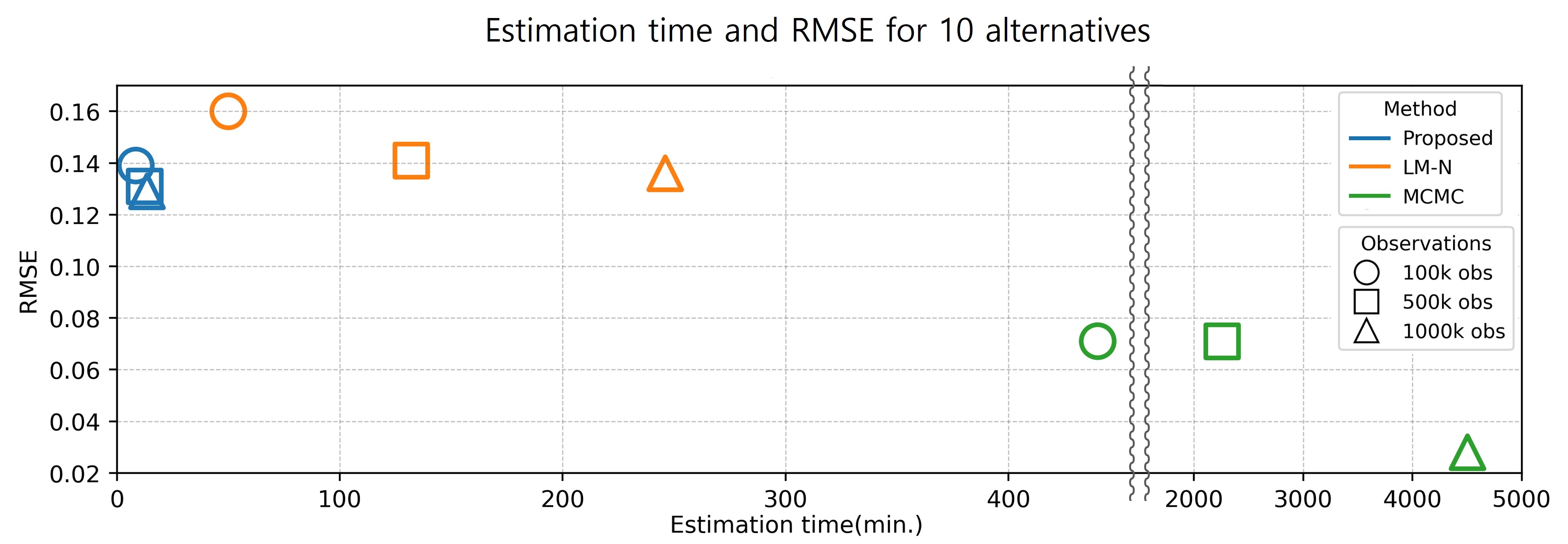}
    \caption{Comparison of estimation time and RMSE for alternatives:10, Observations:1,000,000}
    \label{fig:7a}
\end{subfigure}

\vspace{1em} 

\begin{subfigure}[t]{\columnwidth}
    \centering
    \includegraphics[width=\linewidth]{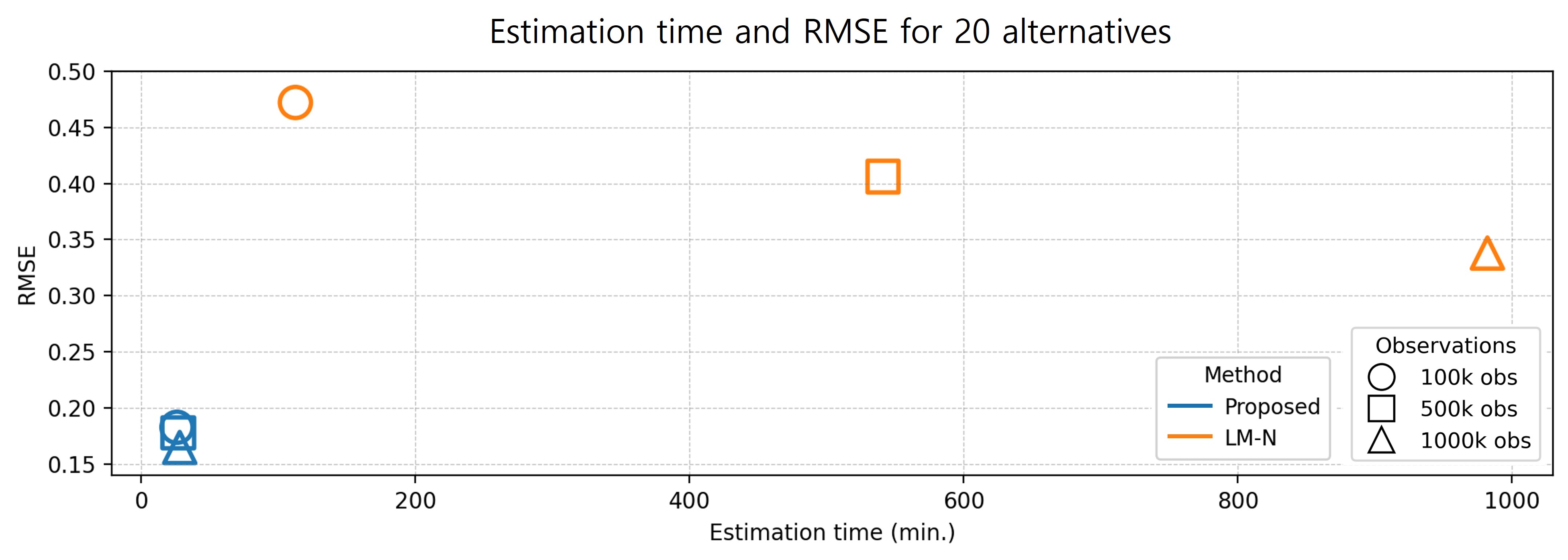}
    \caption{Comparison of estimation time and RMSE for alternatives:20, Observations:1,000,000}
    \label{fig:7b}
\end{subfigure}
\caption{Comparison of estimation time and RMSE in large scale settings}

\end{figure}

Fig. \ref{fig:7a} and \ref{fig:7b} present a more comprehensive comparison of the proposed method, LM–N (\(p = d/2\)) and MCMC in terms of estimation time and RMSE as functions of choice‐set dimension ($d=10$, $d=20$) and observations (100,000, 500,000, and 1,000,000). For 10 alternatives (Fig.\ \ref{fig:7a}), the proposed method takes 8–14 minutes across all observations, whereas LM–N requires 50–250 minutes. MCMC is significantly slower, with the estimation time increasing from about 440 minutes for 100 thousand observations to over 4500 minutes for 1 million observations. For 20 alternatives (Fig.\ \ref{fig:7b}), the proposed method’s estimation time scales almost linearly to 26–28 minutes and remains robust to the change in observations. However, the estimation time of LM–N increases steeply, ranging from 112 to 1,016 minutes, with a significant increase as the number of observations grows.In terms of parameter recovery, the proposed method yields a stable RMSE of 0.12–0.14 at $d=10$ and 0.16–0.19 at $d=20$, largely unaffected by the number of observations. In contrast, LM–N exhibits slight RMSE reduction as observation increases but suffers a significant performance drop when moving from $d=10$(0.13–0.16) to $d=20$(0.34–0.45).

In summary, the estimation time of the proposed method scales almost linearly with the number of alternatives, while that of the LM-N method increases nonlinearly, becoming increasingly impractical as the choice set expands. Unlike LM-N, whose estimation time also grows with the number of observations, the proposed method remains unaffected by sample size. Moreover, while LM-N's parameter recovery performance deteriorates significantly with more alternatives, the proposed method consistently maintains a stable RMSE. 

\begin{figure}[H]
\centering
\begin{subfigure}[t]{0.9\columnwidth}
    \centering
    \includegraphics[width=\columnwidth]{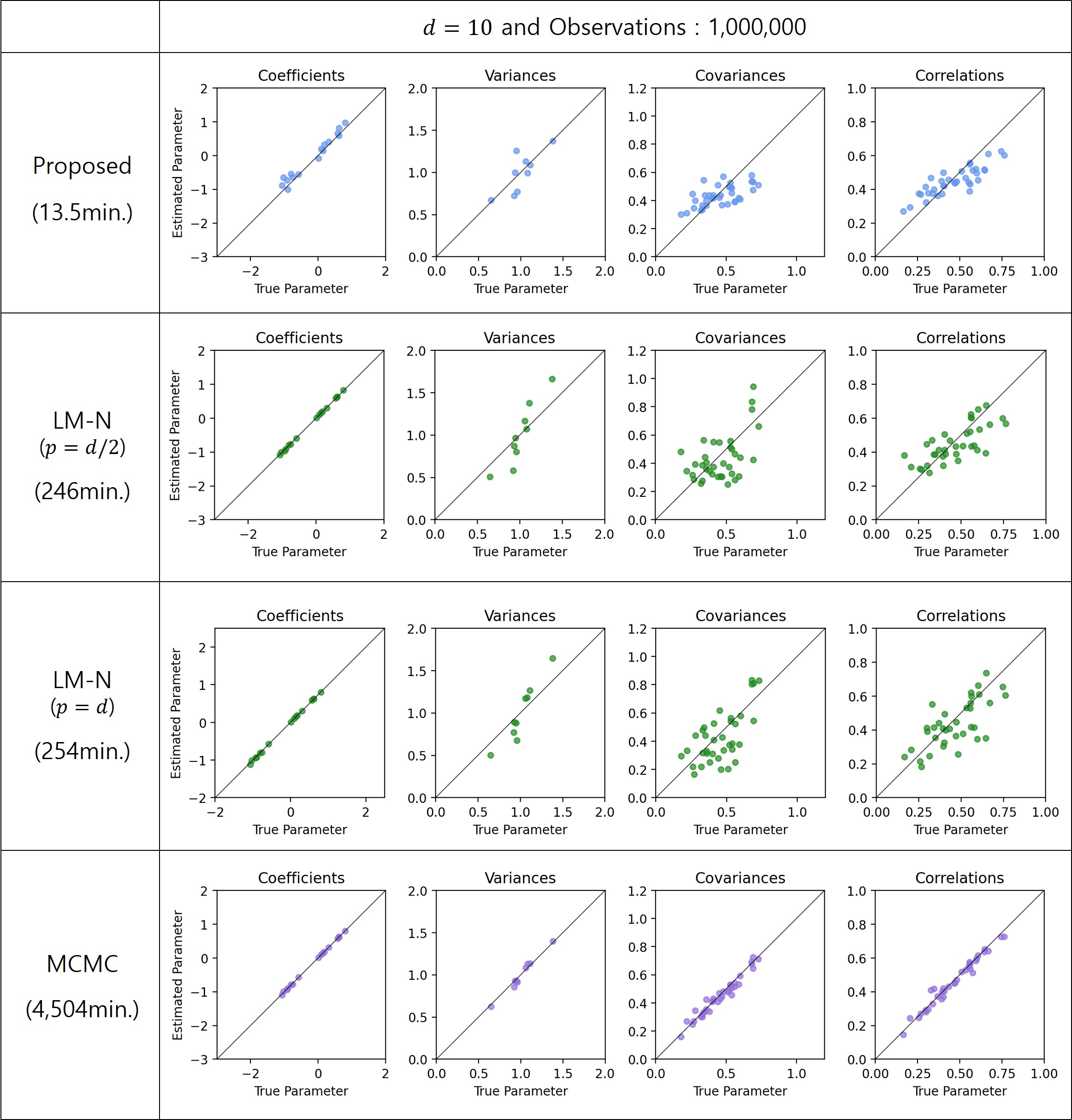}
    \caption{Comparison of true parameters and posterior means(sampling means) for 10 alternatives in large observations}
    \label{fig:8a}
\end{subfigure}

\vspace{1em} 

\begin{subfigure}[t]{0.9\columnwidth}
    \centering
    \includegraphics[width=\columnwidth]{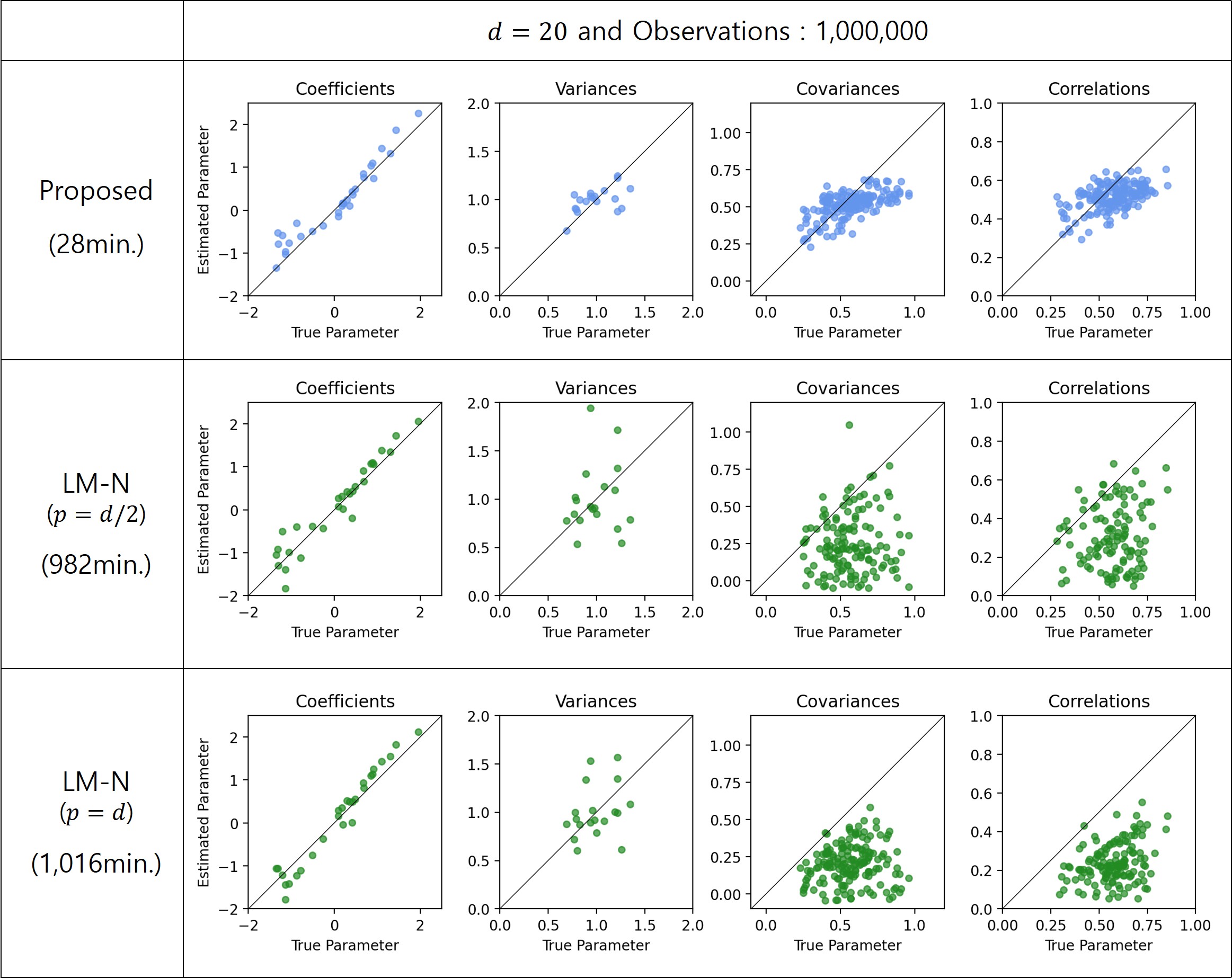}
    \caption{Comparison of true parameters and posterior means(sampling means) for 20 alternatives in large observations}
    \label{fig:8b}
\end{subfigure}
\caption{Comparison of true parameters and posterior means(sampling means) in large scale settings)}
\end{figure}

Fig. \ref{fig:8a} and Figure \ref{fig:8b} present a comparison of true parameters and posterior means of utility parameters, error variances, error covariances, and error correlations for across all methods for simulated data with 1 million observations. MCMC results are excluded for 20 alternatives. Results for 100,000 and 500,000 observations are presented in the Appendix (\ref{appendix:B} and \ref{appendix:C})). 

For 10 alternatives (Fig.\ \ref{fig:8a}), MCMC achieves near-perfect parameter recovery, but its estimation time is impractically long (4,504 minutes). The proposed and LM-N methods recover utility parameters reliably. LM–N’s variance, covariance, and correlation estimates are widely scattered, while the proposed method delivers precise estimates with considerably less bias in covariance and correlation recovery.

As the number of alternatives increases to 20 (Fig.\ \ref{fig:8b}), LM–N’s ability to capture the full variance–covariance structure deteriorates markedly. This decline suggests that the spherical transformation employed in LM-N \citep{bib43} does not scale effectively for large choice sets. Specifically, the method fails to capture inter-dependencies among choice options as dimensionality increases, leading to biases in the covariance estimates. In contrast, the proposed method shows relatively small bias in covariances and correlations estimates, with a pattern similar to that of the ten-alternative scenario. Taste parameter estimates remain tightly clustered around the 45° line, indicating consistent accuracy across large choice‐set sizes. Leveraging a neural embedding framework, it flexibly learns complex relationships among alternatives and thus could recover both utility parameters and error covariance even for large choice‐set sizes.  Overall, these results demonstrate that the proposed method scales robustly in terms of choice‐set dimension and observations, making it particularly well‐suited for multinomial probit estimation in large‐scale applications.

\section{Application to consumer choice data} \label{sec5}
 After establishing the scalability of the proposed method in synthesized datasets, we validated it using real-world data for further deployment. We use the publicly available Dunnhumby dataset titled ``Carbo-Loading: A Relational Database," which contains over one million pasta purchase records across national(US) brands. From this dataset, we used the preprocessed version provided by \cite{bib43}, which excluded private labels and coupon-related transactions. The resulting dataset comprises 10 alternatives and  1,070,436 observations. To ensure consistent comparison with the LM-N method, we implement the same MNP model specification, including an intercept and the log price per ounce for each brand. The utility specification for this study is provided in Appendix \ref{appendix:A}.

Fig.~\ref{fig:9} shows the ten-fold cross-validation results as boxplots of hit rate, log score, and Brier score for the proposed method and LM-N.  For LM-N, we used 10\% sub‐sampling, identical to our large‐scale simulation study, to improve its computational efficiency. 

In terms of estimation time, the proposed method(12.7min.) is roughly 12 times faster than LM-N(152min.) and 344 times faster than MCMC(4,380min.) in the empirical study. In terms of predictive accuracy, the proposed method outperforms LM-N in terms of the hit rate and the Brier score.  Fig.~\ref{fig:9} confirms that these differences are statistically significant. Although the median log score of the proposed method is marginally worse than that of LM-N, this difference could be due to bypassing explicit utility–probability calculations during estimation. The computational efficiency gained by this shortcut is substantial, making the proposed method well-suited for large-scale applications.

\begin{figure}[!ht]
\centering
\includegraphics[width=\columnwidth]{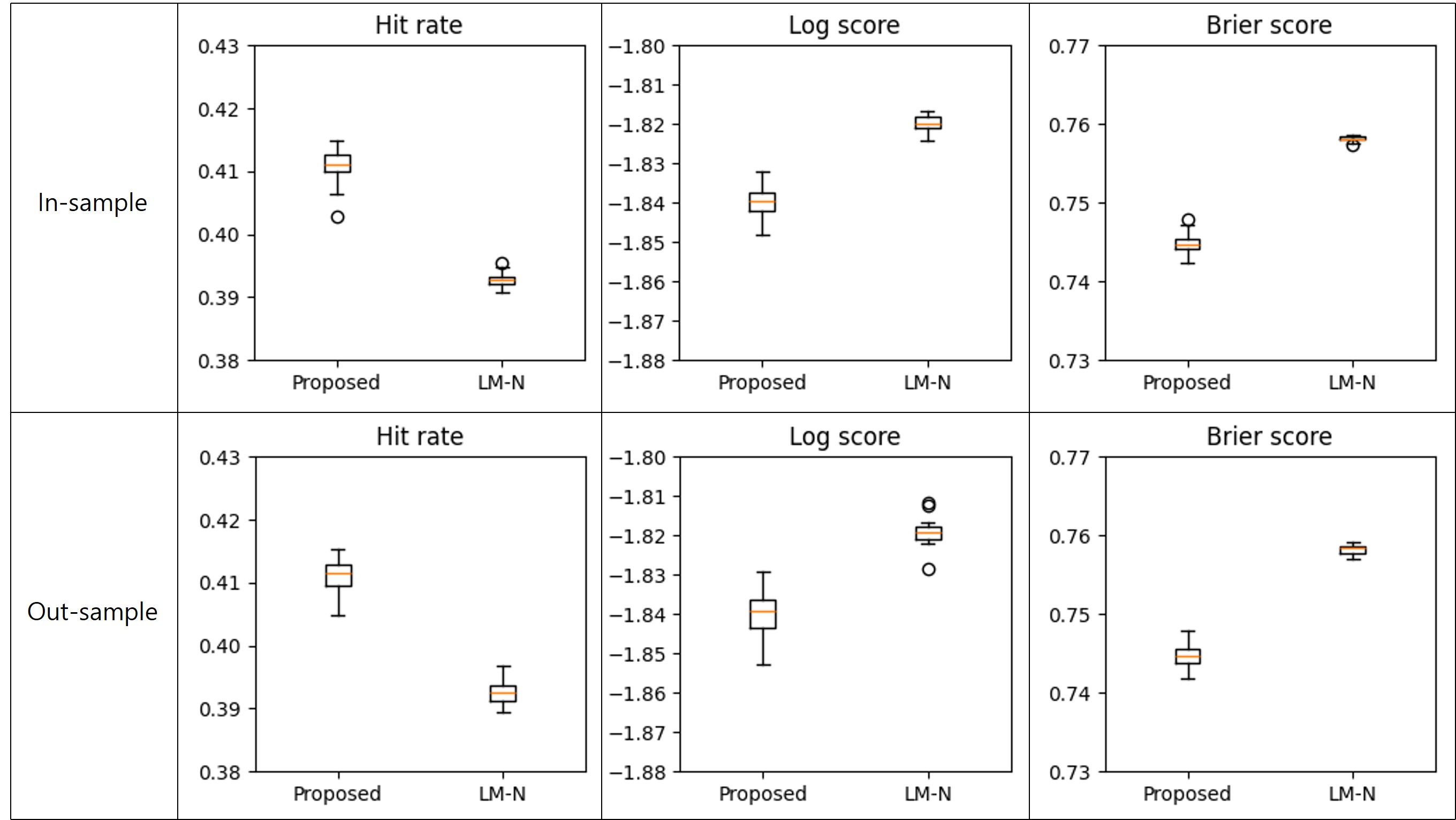}
\caption{Boxplots of hit rate, log score, and brier score for the proposed and LM-N}
\label{fig:9}
\end{figure}

\begin{figure}[!ht]
\centering
\includegraphics[width=\columnwidth]{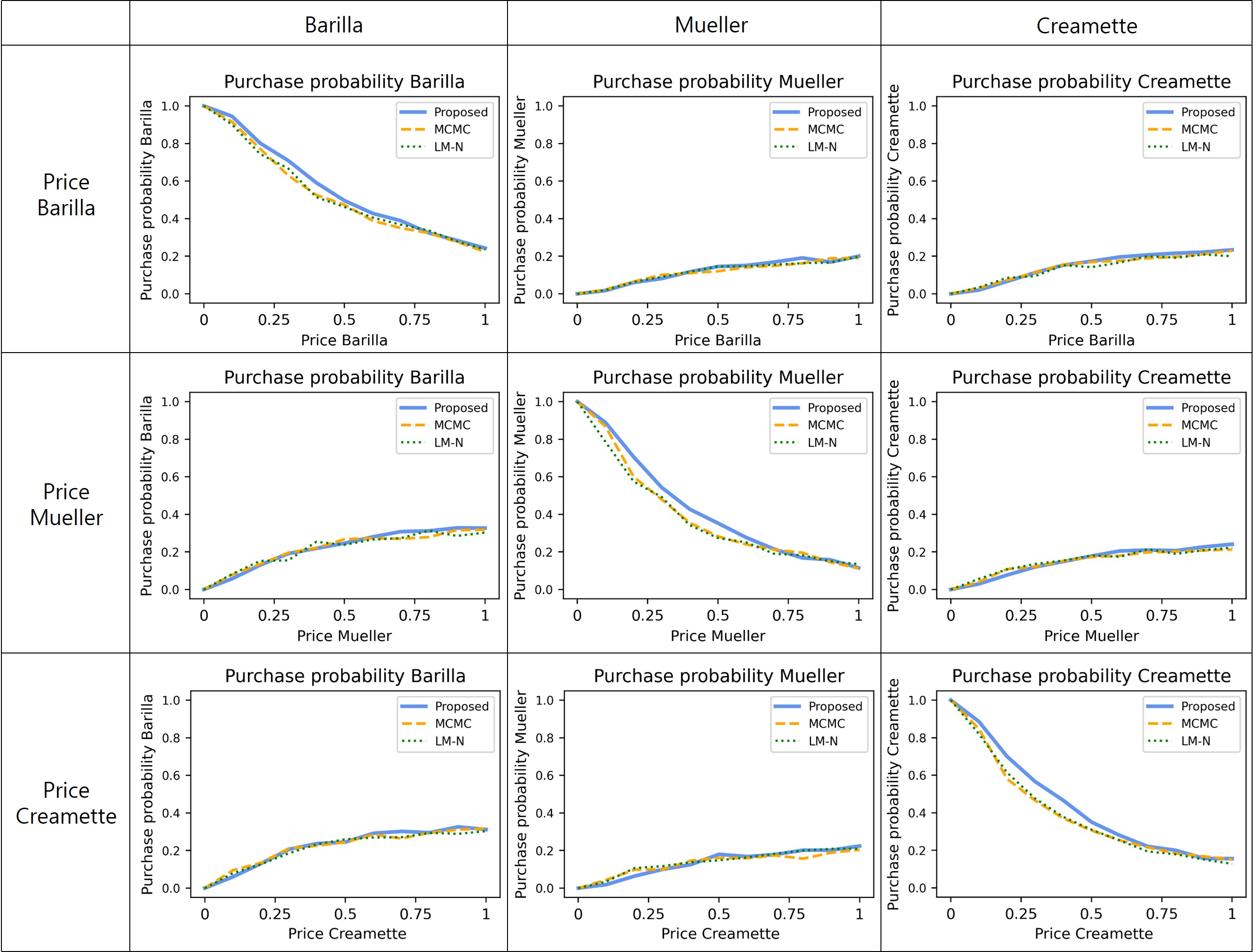}
\caption{Purchase probabilities as a function of pasta price. The solid blue line denotes the proposed method, the dashed orange line denotes MCMC, and the dotted green line denotes LM-N.}
\label{fig:10}
\end{figure}

 Fig.~\ref{fig:10} presents how purchase probabilities change with each brand’s price while keeping the prices of the remaining brands fixed at their mean values. As the pattern is consistent across brands, we only present results for Barilla, Mueller, and Creamette. %
 Rows correspond to varying the price, while columns show the resulting purchase probabilities. Diagonal panels depict own‐price effects, whereas off‐diagonal panels illustrate cross‐price effects, i.e., how a price change in one brand influences the purchase probabilities of the others. When a brand’s price increases and its purchase probability decreases, the purchase probabilities of the other brands increase correspondingly. This pattern is consistent across all three approaches, and they all produce nearly identical probability curves.

Overall, empirical results demonstrate that the proposed method scales effectively to real‐world choice settings with 10 alternatives and over one million observations, maintaining both computational efficiency and predictive accuracy. Its performance and speed in large‐scale empirical data suggest broad applicability in practical discrete choice modeling tasks.

\section{Conclusions}\label{sec6}
This study introduces a scalable and flexible conditional variational inference (CVI) framework for parameter calibration in the multinomial probit (MNP) model, addressing longstanding computational bottlenecks in high-dimensional discrete choice modeling. By reformulating the posterior approximation task using conditional neural embeddings and bypassing intractable integrals via differentiable `argmax' approximations using straight-through estimators and the Gumbel-Softmax trick, the proposed method achieves substantial computational gains without sacrificing estimation accuracy or predictive performance in large-scale datasets.

The CVI approach addresses key challenges in applying variational inference to MNP by ensuring the positive definiteness and identifiability of the latent utility covariance matrix while capturing complex posterior dependencies. Specifically, our framework leverages a reparameterized covariance structure with trace normalization to guarantee identifiability. This formulation enables efficient optimization using stochastic gradient descent, and is compatible with modern automatic differentiation toolkits.

Comprehensive simulation experiments and empirical validation using over one million real-world observations demonstrate the scalability, accuracy, and robustness of the proposed approach. In head-to-head comparisons, our method matches or outperforms state-of-the-art alternatives such as MCMC and the LM-N variational approach in terms of parameter recovery, predictive performance, and computational efficiency. Notably, it achieves up to 36× faster estimation than LM-N and remains stable in parameter recovery even with 20 alternatives and one million observations, a scale at which MCMC becomes computationally infeasible.

While the proposed method offers point estimates rather than full posterior distributions, we show that bootstrap-based inference provides reliable uncertainty quantification. 
Overall, this work expands the toolbox for scalable discrete choice modeling by bridging statistical and machine learning methods. The proposed CVI framework holds promise not only for large-scale MNP estimation but also for other high-dimensional latent variable models where exact posterior inference is intractable. Future research could extend the framework to accommodate dynamic choice behavior models and panel data structures. \label{sec7}

\bibliography{sn-bibliography}

\section{Acknowledgement}
This research was supported by the National Research Foundation of Korea (NRF) grant funded by the Korean government (RS-2024-00337956) and Korea Agency for Infrastructure Technology Advancement (KAIA) grant funded by the Ministry of Land, Infrastructure and Transport (RS-2022-00141102). and the Chung-Ang University Graduate Research Scholarship in 2022

\clearpage
\appendix

\section*{Appendix}  
\section{Simulation and empirical studies settings}\label{appendix:A}

\subsection{Hyper-parameters used in simulation and empirical studies}
\begin{table}[!ht]
  \centering
  \label{tab:study-settings}
  \scriptsize
  \setlength\tabcolsep{6pt}
  \renewcommand{\arraystretch}{1.1}

  \begin{tabular}{@{}
      >{\centering\arraybackslash}m{2cm}
      >{\centering\arraybackslash}m{3cm}
      c c c c
    @{}}
    \toprule
    \makecell{Study}
      & \makecell{Scale}
      & \makecell{Learning rate\\($\alpha$)}
      & \makecell{Temperature\\($\beta$)}
      & \makecell{Minibatch size\\($m$)}
      & \makecell{Number of utility\\samples ($L$)} \\
    \midrule

    \multirow{2}{*}{\adjustbox{valign=c}{\shortstack{Simulation\\study}}}
      & \shortstack{Small scale\\($d=3$)}
      & \multirow{2}{*}{\adjustbox{valign=c}{$0.001$}}
      & \multirow{2}{*}{\adjustbox{valign=c}{$0.1\sim0.01$}}
      & \multirow{2}{*}{\adjustbox{valign=c}{$500$}}
      & 20 \\

      & \shortstack{Large scale\\($d=10$ and $d=20$)}
      & 
      & 
      & 
      & 100 \\

    \midrule
    Empirical study
      & $d=10$
      & 0.001
      & $0.1\sim0.01$
      & 500
      & 100 \\
    \bottomrule
  \end{tabular}
\end{table}

\subsection{MNP model in empirical study}
In the empirical study, the choice set has 10 pasta brands (alternatives): 
Barilla, Mueller, Creamette, Ronzoni, San~Giorgio, No~Yolks, Hodgson~Mills, Healthy~Harvest, DaVinci, DeCecco. Let \(X_{ij}\) be the log(price) per ounce of brand \(j\) in observation \(i\). The $(10 \times 10)$ identity matrix encompasses dummy variables for the alternative‐specific intercepts of each brand. The design and parameter matrices of the MNP model are as follows:

\[
X_i
= \left[
  \begin{pmatrix}
1 & 0 & 0 & \cdots & 0 \\
0 & 1 & 0 & \cdots & 0 \\
0 & 0 & 1 & \cdots & 0 \\
\vdots & \vdots & \vdots & \ddots & \vdots \\
0 & 0 & 0 & \cdots & 1
\end{pmatrix}_{10 \times 10}\;\Big|\;
     \begin{pmatrix}
       X_{i,\mathrm{Barilla}}\\[3pt]
       X_{i,\mathrm{Mueller}}\\[3pt]
       X_{i,\mathrm{Creamette}}\\[3pt]
       X_{i,\mathrm{Ronzoni}}\\[3pt]
       X_{i,\mathrm{San\,Giorgio}}\\[3pt]
       X_{i,\mathrm{No\,Yolks}}\\[3pt]
       X_{i,\mathrm{Hodgson\,Mills}}\\[3pt]
       X_{i,\mathrm{Healthy\,Harvest}}\\[3pt]
       X_{i,\mathrm{DaVinci}}\\[3pt]
       X_{i,\mathrm{DeCecco}}
     \end{pmatrix}
\right]_{10\times11} and
\]

\[
\boldsymbol{a}
=
\begin{pmatrix}
\alpha_1 \\ \alpha_2 \\ \vdots \\ \alpha_{10} \\[4pt]
\gamma
\end{pmatrix}_{11\times 1},
\]
where we set \(\alpha_1=0\) (Barilla as reference) for identification, \(\alpha_2,\dots,\alpha_{10}\) are alternative‐specific intercepts, and \(\gamma\) is the generic parameter (log(price)‐sensitivity) applied to all brands.

\clearpage 
\section{Comparison of the proposed method against LM-N and MCMC across 10 alternatives}\label{appendix:B}

\subsection{Comparison of the proposed method against LM-N and MCMC}
\begin{table}[h]
\centering
\label{tab:ap1}
\renewcommand{\arraystretch}{1.3}
\begin{tabular}{@{}c c c c c c c c@{}}
\toprule
\textbf{Obs.} & \textbf{Model} & \textbf{Sample} & \textbf{Hit rate} & \textbf{Log score} & \textbf{Brier score} & \textbf{RMSE} & \textbf{\begin{tabular}[c]{@{}c@{}}Est. time\\ (min.)\end{tabular}} \\
\midrule
\multirow{8}{*}{100{,}000} 
  & \multirow{2}{*}{Proposed method} & In-sample   & 0.252 & -1.968 & 0.836 & \multirow{2}{*}{0.139} & \multirow{2}{*}{8.5}   \\
  &                                 & Out  & 0.252 & -1.988 & 0.837 &                        &                       \\
  \cmidrule(lr){2-8}
  & \multirow{2}{*}{LM-N($p=d/2$)}         & In-sample   & 0.252 & -1.971 & 0.834 & \multirow{2}{*}{0.160} & \multirow{2}{*}{50}    \\
  &                                 & Out-of-sample
  & 0.250 & -1.979 & 0.836 &                        &                       \\
  \cmidrule(lr){2-8}
  & \multirow{2}{*}{LM-N($p=d$)}        & In-sample   & 0.251 & -2.027 & 0.843 & \multirow{2}{*}{0.307} & \multirow{2}{*}{57}    \\
  &                                 & Out-of-sample
  & 0.249 & -2.030 & 0.844 &                        &                       \\
  \cmidrule(lr){2-8}
  & \multirow{2}{*}{MCMC}            & In-sample   & 0.254 & -1.960 & 0.835 & \multirow{2}{*}{0.071} & \multirow{2}{*}{440}   \\
  &                                 & Out-of-sample
  & 0.253 & -1.965 & 0.836 &                        &                       \\
\midrule
\multirow{8}{*}{500{,}000}
  & \multirow{2}{*}{Proposed method} & In-sample   & 0.253 & -1.981 & 0.836 & \multirow{2}{*}{0.131} & \multirow{2}{*}{12.4}  \\
  &                                 & Out-of-sample
  & 0.251 & -1.981 & 0.837 &                        &                       \\
  \cmidrule(lr){2-8}
  & \multirow{2}{*}{LM-N($p=d/2$)}         & In-sample   & 0.254 & -1.971 & 0.834 & \multirow{2}{*}{0.141} & \multirow{2}{*}{132}   \\
  &                                 & Out-of-sample
  & 0.248 & -1.978 & 0.834 &                        &                       \\
  \cmidrule(lr){2-8}
  & \multirow{2}{*}{LM-N($p=d$)}        & In-sample   & 0.253 & -1.972 & 0.834 & \multirow{2}{*}{0.134} & \multirow{2}{*}{144}   \\
  &                                 & Out-of-sample
  & 0.248 & -1.977 & 0.835 &                        &                       \\
  \cmidrule(lr){2-8}
  & \multirow{2}{*}{MCMC}            & In-sample   & 0.256 & -1.958 & 0.834 & \multirow{2}{*}{0.071} & \multirow{2}{*}{2256} \\
  &                                 & Out-of-sample
  & 0.255 & -1.964 & 0.836 &                        &                       \\
\midrule
\multirow{8}{*}{1{,}000{,}000}
  & \multirow{2}{*}{Proposed method} & In-sample   & 0.255 & -1.971 & 0.834 & \multirow{2}{*}{0.129} & \multirow{2}{*}{13.5}  \\
  &                                 & Out-of-sample
  & 0.251 & -1.986 & 0.836 &                        &                       \\
  \cmidrule(lr){2-8}
  & \multirow{2}{*}{LM-N($p=d/2$)}         & In-sample   & 0.254 & -1.969 & 0.834 & \multirow{2}{*}{0.136} & \multirow{2}{*}{246}   \\
  &                                 & Out-of-sample
  & 0.251 & -1.971 & 0.834 &                        &                       \\
  \cmidrule(lr){2-8}
  & \multirow{2}{*}{LM-N($p=d$)}        & In-sample   & 0.254 & -1.972 & 0.834 & \multirow{2}{*}{0.131} & \multirow{2}{*}{254}   \\
  &                                 & Out-of-sample
  & 0.250 & -1.974 & 0.834 &                        &                       \\
  \cmidrule(lr){2-8}
  & \multirow{2}{*}{MCMC}            & In-sample   & 0.257 & -1.954 & 0.834 & \multirow{2}{*}{0.028} & \multirow{2}{*}{4504} \\
  &                                 & Out-of-sample
  & 0.255 & -1.958 & 0.836&                        &                       \\
\bottomrule
\end{tabular}
\end{table}

\FloatBarrier

\clearpage 
\subsection{Comparison of true parameters and posterior means(sampling means) in 100,000 observations}
\begin{figure}[!ht]
\centering
\includegraphics[width=\columnwidth]{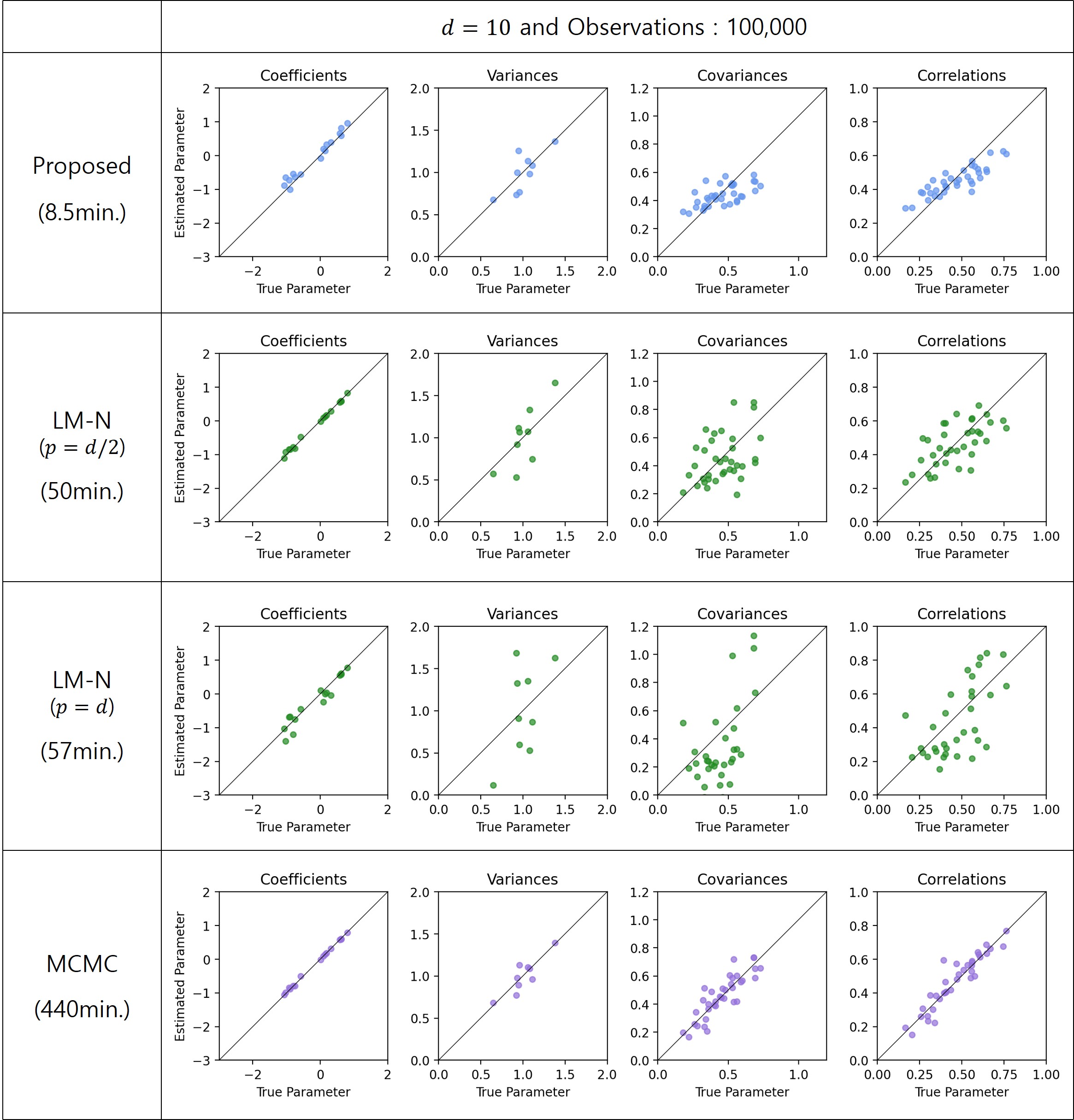}
\label{fig:B2}
\end{figure}
\FloatBarrier

\clearpage 
\subsection{Comparison of true parameters and posterior means(sampling means) in 500,000 observations}
\begin{figure}[!ht]
\centering
\includegraphics[width=\columnwidth]{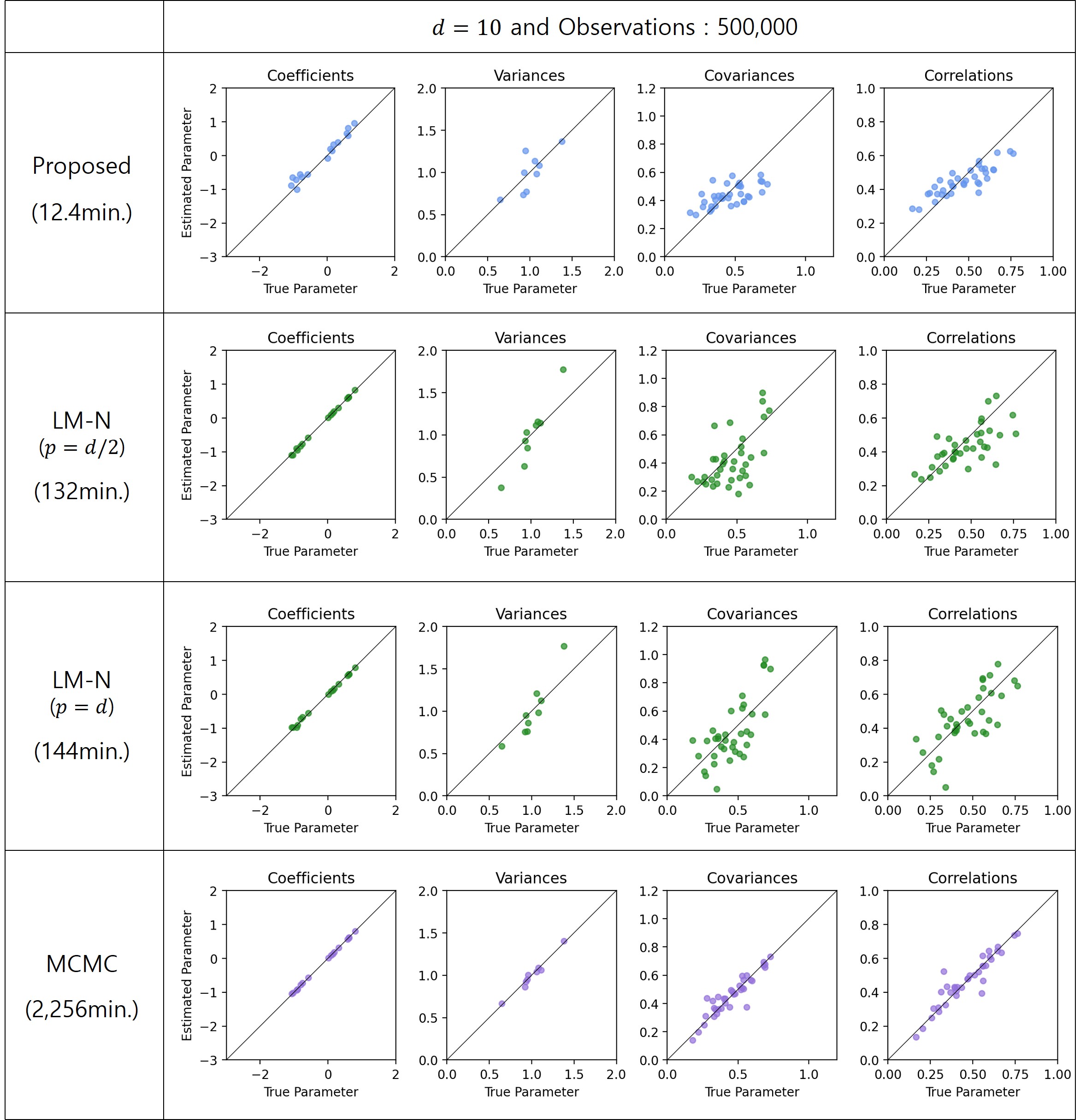}
\label{fig:B3}
\end{figure}
\FloatBarrier

\clearpage 
\subsection{Comparison of true parameters and posterior means(sampling means) in 1,000,000 observations}
\begin{figure}[!ht]
\centering
\includegraphics[width=\columnwidth]{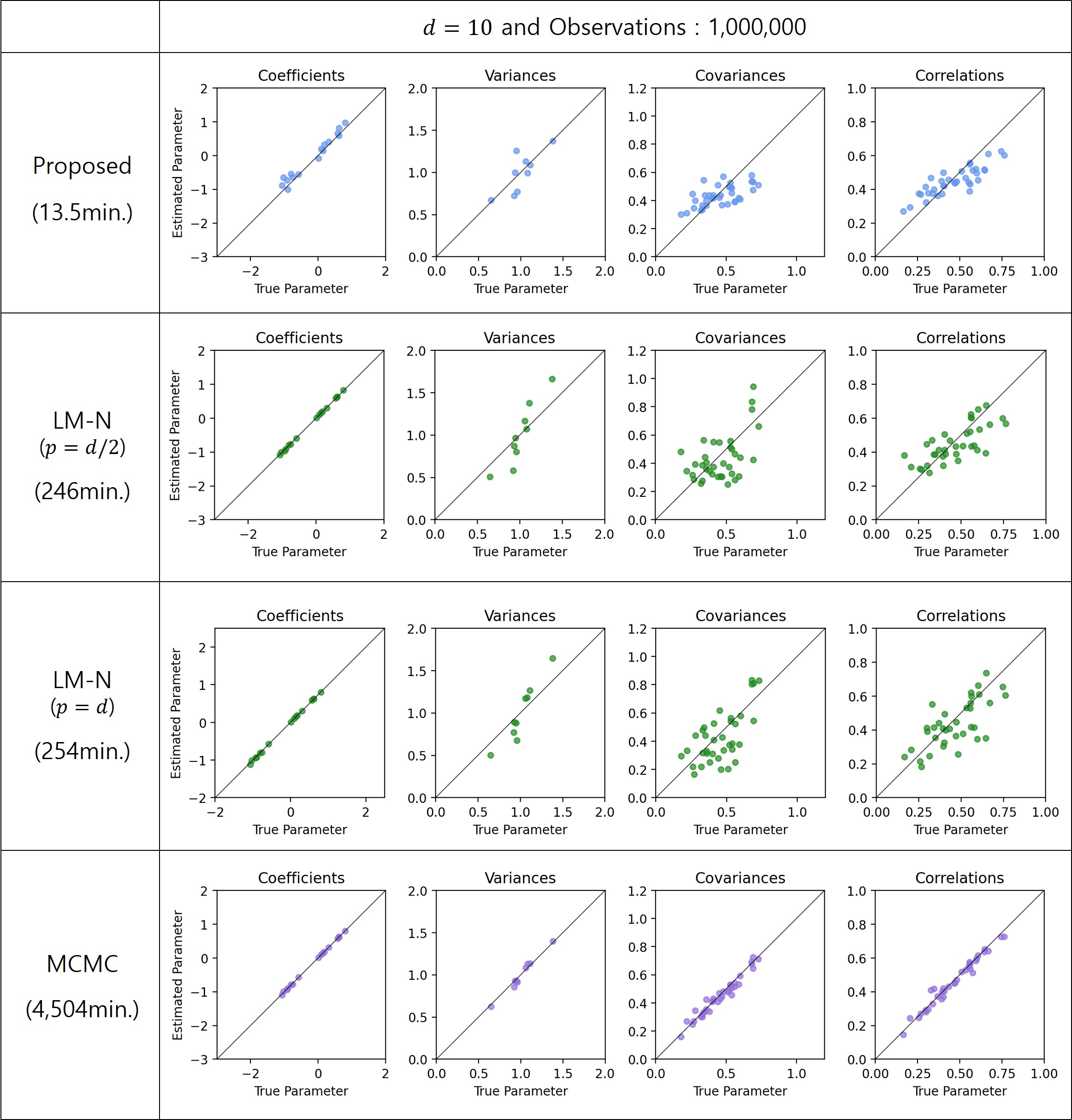}
\label{fig:B4}
\end{figure}
\FloatBarrier

\clearpage 
\section{Comparison of the proposed method against LM-N across 20 alternatives}\label{appendix:C}

\subsection{Comparison of the proposed method against LM-N}
\begin{table}[h]
\centering
\label{tab:ap2}
\renewcommand{\arraystretch}{1.3}
\begin{tabular}{@{}c c c c c c c c@{}}
\toprule
\textbf{Obs.} & \textbf{Model} & \textbf{Sample} & \textbf{Hit rate} & \textbf{Log score} & \textbf{Brier score} & \textbf{RMSE} & \textbf{\begin{tabular}[c]{@{}c@{}}Est. time\\ (min.)\end{tabular}} \\
\midrule
\multirow{6}{*}{100{,}000}
  & \multirow{2}{*}{Proposed method} & In-sample  & 0.459 & -1.902 & 0.701 & \multirow{2}{*}{0.183} & \multirow{2}{*}{26}   \\
  &                                 & Out-of-sample & 0.458 & -1.904 & 0.702 &                        &                        \\
  \cmidrule(lr){2-8}
  & \multirow{2}{*}{LM-N($p=d/2$)}        & In-sample  & 0.452 & -1.907 & 0.702 & \multirow{2}{*}{0.465} & \multirow{2}{*}{112} \\
  &                                 & Out-of-sample & 0.452 & -1.915 & 0.704 &                        &                        \\
  \cmidrule(lr){2-8}
  & \multirow{2}{*}{LM-N($p=d$)}        & In-sample  & 0.452 & -1.898 & 0.701 & \multirow{2}{*}{0.473} & \multirow{2}{*}{165} \\
  &                                 & Out-of-sample & 0.452 & -1.907 & 0.705 &                        &                        \\
\midrule
\multirow{6}{*}{500{,}000}
  & \multirow{2}{*}{Proposed method} & In-sample  & 0.459 & -1.887 & 0.701 & \multirow{2}{*}{0.178} & \multirow{2}{*}{27}   \\
  &                                 & Out-of-sample & 0.459 & -1.899 & 0.702 &                        &                        \\
  \cmidrule(lr){2-8}
  & \multirow{2}{*}{LM-N($p=d/2$)}        & In-sample  & 0.452 & -1.881 & 0.701 & \multirow{2}{*}{0.406} & \multirow{2}{*}{541} \\
  &                                 & Out-of-sample & 0.454 & -1.902 & 0.705 &                        &                        \\
  \cmidrule(lr){2-8}
  & \multirow{2}{*}{LM-N($p=d$)}        & In-sample  & 0.453 & -1.902 & 0.701 & \multirow{2}{*}{0.472} & \multirow{2}{*}{588} \\
  &                                 & Out-of-sample & 0.452 & -1.907 & 0.705 &                        &                        \\
\midrule
\multirow{6}{*}{1{,}000{,}000}
  & \multirow{2}{*}{Proposed method} & In-sample  & 0.461 & -1.897 & 0.701 & \multirow{2}{*}{0.165} & \multirow{2}{*}{28}   \\
  &                                 & Out-of-sample & 0.460 & -1.889 & 0.702 &                        &                        \\
  \cmidrule(lr){2-8}
  & \multirow{2}{*}{LM-N($p=d/2$)}        & In-sample  & 0.454 & -1.874 & 0.701 & \multirow{2}{*}{0.344} & \multirow{2}{*}{982} \\
  &                                 & Out-of-sample & 0.452 & -1.891 & 0.704 &                        &                        \\
  \cmidrule(lr){2-8}
  & \multirow{2}{*}{LM-N($p=d$)}        & In-sample  & 0.452 & -1.896 & 0.701 & \multirow{2}{*}{0.447} & \multirow{2}{*}{1016}\\
  &                                 & Out-of-sample & 0.451 & -1.898 & 0.704 &                        &                        \\
\bottomrule
\end{tabular}
\end{table}

\FloatBarrier

\clearpage 
\subsection{Comparison of true parameters and posterior means(sampling means) in 100,000 observations}
\centering
\begin{figure}[!htbp]
\centering
\includegraphics[width=\columnwidth]{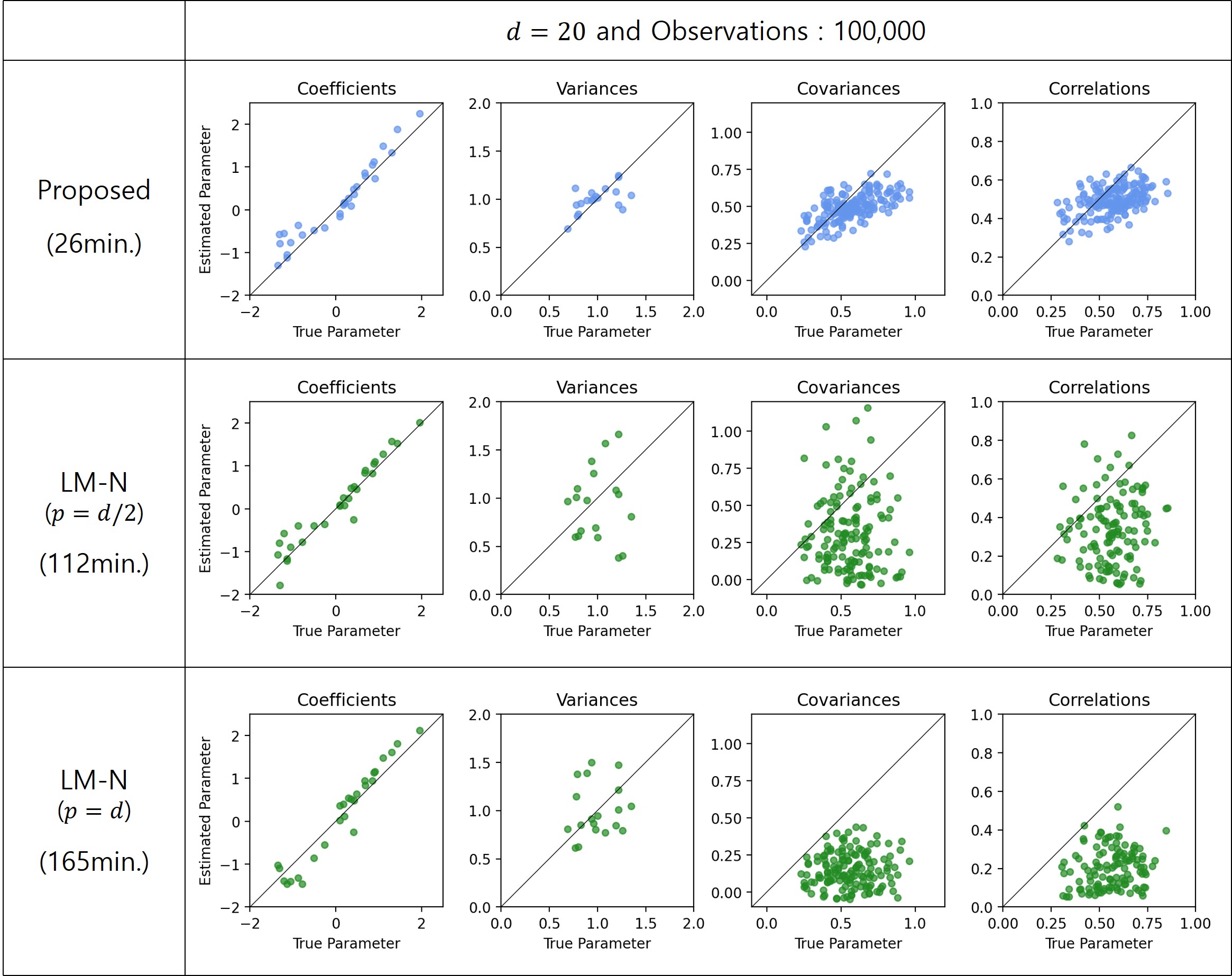}
\label{fig:C2}
\end{figure}

\clearpage 
\subsection{ Comparison of true parameters and posterior means(sampling means) in 500,000 observations}
\begin{figure}[!htbp]
\centering
\includegraphics[width=\columnwidth]{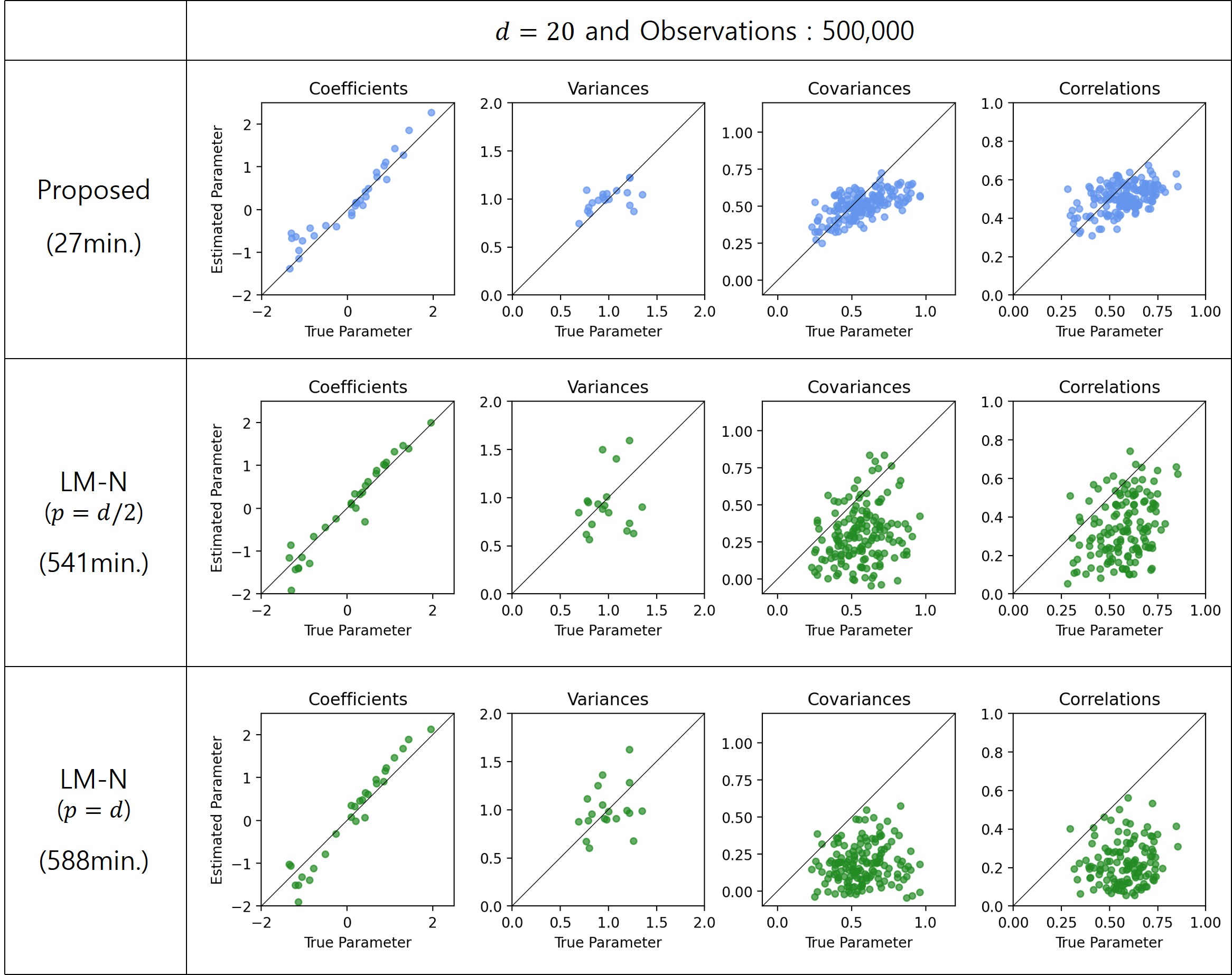}
\label{fig:C3}
\end{figure}
\FloatBarrier

\clearpage 
\subsection{ Comparison of true parameters and posterior means(sampling means) in 1,000,000 observations}
\begin{figure}[!ht]
\centering
\includegraphics[width=\columnwidth]{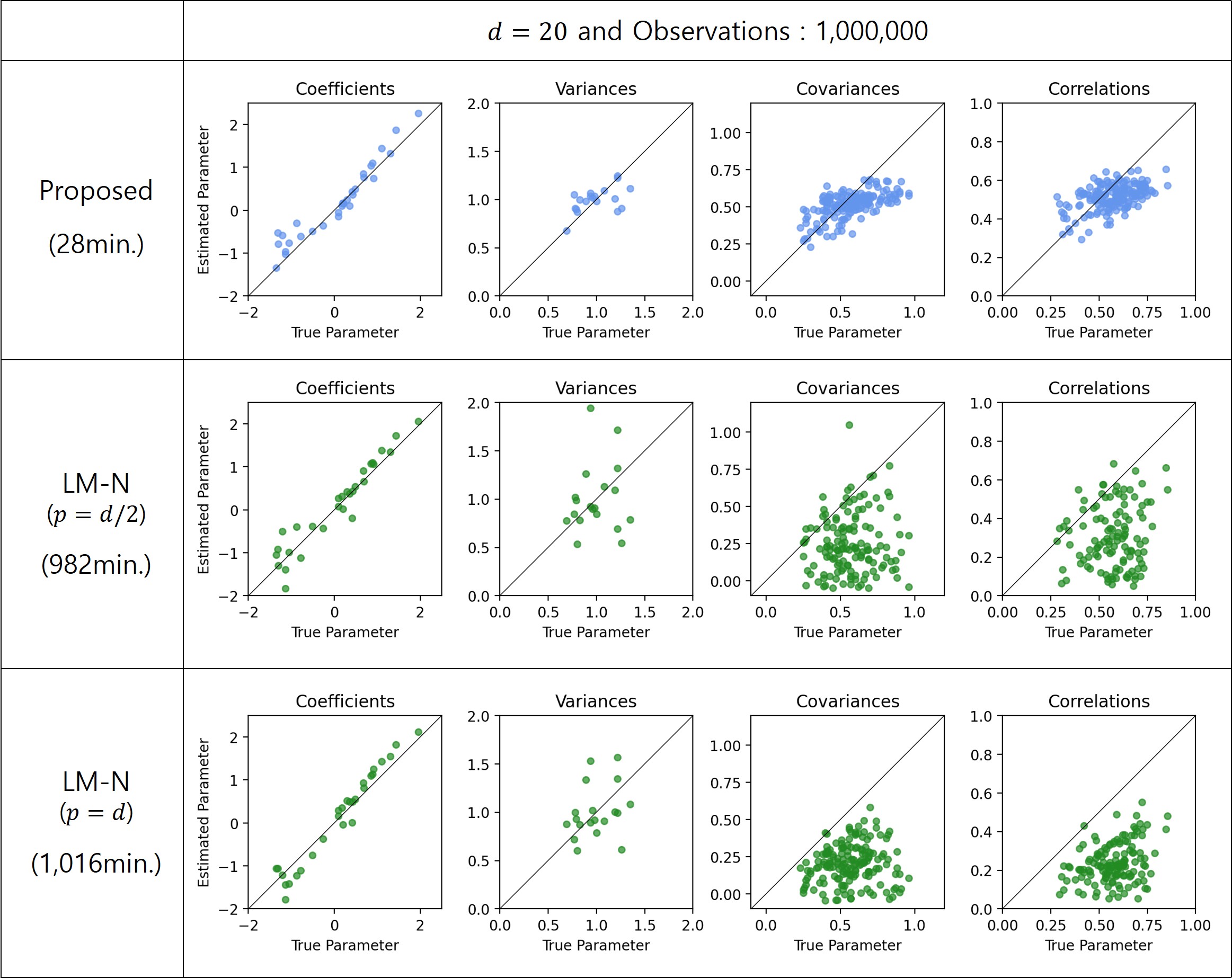}
\label{fig:C4}
\end{figure}
\FloatBarrier

\end{document}